\pdfoutput=1

\documentclass[11pt,twoside,a4paper,cmspaper,final,collab]{cms-tdr}
\def\svnVersion{208714}\def\cmsCernNo{2012-077}\def\cmsCernDate{\today}\def\cmsMessage{Published in the Journal of High Energy Physics as \href{http://dx.doi.org/10.1007/JHEP09(2012)029}{\doi{10.1007/JHEP09(2012)029.}}}

\begin{document}\cmsNoteHeader{EXO-11-006}

\hyphenation{had-ron-i-za-tion}
\hyphenation{cal-or-i-me-ter}
\hyphenation{de-vices}

\RCS$Revision: 141324 $
\RCS$HeadURL: svn+ssh://alverson@svn.cern.ch/reps/tdr2/papers/EXO-11-006/trunk/EXO-11-006.tex $
\RCS$Id: EXO-11-006.tex 141324 2012-07-31 13:59:01Z srappocc $
\providecommand{\rd}{\ensuremath{\mathrm{d}}} 
\cmsNoteHeader{EXO-11-006} 
\title{Search for anomalous $\ttbar$ production in the highly-boosted all-hadronic final state}

\author{The CMS Collaboration}

\newcommand{\PYTHIAEIGHT} {{\textsc{pythia8}}\xspace}
\newcommand{\pdf}{\ensuremath{\mathrm{pdf}}}
\newcommand{\ta}{\mbox{$\tau$}}
\newcommand{\mtt}{\ensuremath{m_{\mathrm{t}\overline{\mathrm{t}}}}~}
\newcommand{\wboson}{\ensuremath{{\mathrm{W}}}}
\newcommand{\zboson}{\ensuremath{{\mathrm{Z}}}}
\newcommand{\mzp}{\ensuremath{m_{\mathrm{Z^{\prime}}}}}
\newcommand{\qq}{$\mathrm{q}\bar{\mathrm{q}}$~}
\newcommand{\invpb}{pb$^{-1}$~}
\newcommand{\intlumi}{5\fbinv}
\newcommand{\wmasssemilepdata}{ \ensuremath{m_\wboson^{\text{DATA}} = 83.0 \pm 0.7\GeVcc}}
\newcommand{\wmasssemilepmc}{ \ensuremath{m_\wboson^{\mathrm{MC}} = 82.5 \pm 0.3\GeVcc}}
\newcommand{\topmasssemilepdata}{ \ensuremath{m_{\mathrm{t}}^{\text{DATA}} = 177.1 \pm 2.0\GeVcc}}
\newcommand{\topmasssemilepmc}{ \ensuremath{m_{\mathrm{t}}^{\mathrm{MC}} = 171.5 \pm 0.7\GeVcc}}
\newcommand{\subjetjes}{\ensuremath{1.01 \pm 0.01}}
\newcommand{\mcuteffsemilepdata}{ \ensuremath{\epsilon_{m_\wboson}^{\text{DATA}} = 0.49 \pm 0.01}}
\newcommand{\mcuteffsemilepmc}{ \ensuremath{\epsilon_{m_\wboson}^{\mathrm{MC}} =  0.50 \pm 0.01}}
\newcommand{\mucuteffsemilepdata}{ \ensuremath{\epsilon_\mu^{\text{DATA}} = 0.64 \pm 0.01}}
\newcommand{\mucuteffsemilepmc}{ \ensuremath{\epsilon_\mu^{\mathrm{MC}} =  0.64 \pm 0.01}}
\newcommand{\scalefactor}{ \ensuremath{0.97 \pm 0.03}}
\newcommand{\scalefactorsqrd}{ \ensuremath{0.94 \pm 0.06}}
\newcommand{\topmasscuteffsemilepdata}{ \ensuremath{\epsilon_{m_t}^{\text{DATA}} = 0.50 \pm 0.08}}
\newcommand{\topmasscuteffsemilepmc}{ \ensuremath{\epsilon_{m_t}^{\mathrm{MC}} =  0.65}}
\newcommand{\minmasscuteffsemilepdata}{ \ensuremath{\epsilon_\text{minMass}^{\text{DATA}} = 0.71 \pm 0.16}}
\newcommand{\minmasscuteffsemilepmc}{ \ensuremath{\epsilon_\text{minMass}^{\mathrm{MC}} =  0.62}}
\newcommand{\scalefactorcatoptag}{ \ensuremath{0.89 \pm 0.28}}

\newcommand{\ifnpas}{\iffalse}
\newcommand{\ifpas}{\iftrue}

\date{\today}

\abstract{
A search is presented for a massive particle,
generically referred to as a $\zp$, decaying
into a $\ttbar$ pair.  The search focuses on $\zp$ resonances
that are sufficiently massive to produce highly Lorentz-boosted
top quarks,
which yield collimated decay products that are partially or fully merged into single jets.
The analysis uses new methods to analyze jet substructure, providing
suppression of the
non-top multijet
backgrounds.
The analysis is based on a data sample of proton-proton collisions at a center-of-mass energy of 7\TeV,
corresponding to an integrated luminosity of \intlumi.
Upper limits in the range of 1\,pb are set on the product of
the production
cross section and branching fraction for a topcolor $\zp$ modeled
for
several widths,
as well as for a
Randall--Sundrum Kaluza--Klein gluon.
In addition, the results constrain any enhancement in $\ttbar$
production beyond expectations of the standard model for $\ttbar$
invariant mass larger than 1 \TeVcc.
}

\hypersetup{%
pdfauthor={CMS Collaboration},%
pdftitle={Search for anomalous t t-bar production in the highly-boosted all-hadronic final state},%
pdfsubject={CMS},%
pdfkeywords={CMS, physics, software, computing}}

\maketitle 

\ifnpas

{\bf Changes since the preapproval}
\begin{itemize}
\item Updated the RS KKG model to 7 TeV via private communication with
  the authors.
\item Was not correctly taking into account the ``substructure scale
  factor'' in the analysis (was doing $SF$, not $SF^2$ because we're
  using double tags).
\item Now weighting the leading jet \pt by the trigger turnon
  curve derived from the Monte Carlo.
\item Now removing \ttbar from the mistag rate estimate.
\item Found a bug in the mistag rate derivation procedure for the 1+1
  case, was only taking the subleading jet by mistake (should be
  randomly picking leading and subleading).
\item The JES uncertainty for the \ttbar background was incorrectly
  set to the JES uncertainty from \zp of 1 TeV.
\item Restructured the AN and PAS a little so that we don't ``violate
  causality'' for the flow of the paper.
\end{itemize}

{\bf Changes since publication of EXO-11-006 PAS}
\begin{itemize}
\item Updated to \intlumi. Will update to 5 in the immediate future.
\item Switched the ``mass-modified-mistag'' procedure to use a prior
  from QCD MC instead of a flat prior on the jet mass for the probe
  jet. The same procedure is used for the systematic uncertainty (half
  the difference between ``mass-modified'' and no correction).
\item Updated scale factor measurement to use \intlumi.
\item Added an analysis to limit the possible enhancements to the
  $\ttbar$ production cross section that could be present.
\item Added a shape analysis.
\item Changed the sideband definition for the ``1+2'' analysis, now
  uses the top-tagging rate derived from a sample with the $W$-tagging
  ``$\mu$ cut'' reversed. This removed a discrepancy of around 10\%.
\end{itemize}

\fi

\section{Introduction}

\ifnpas
\section{Introduction}
\fi
\label{sec:introduction}

Among scenarios for physics beyond the standard model (SM) are possibilities of new
gauge interactions with large couplings to third-generation
quarks \cite{mssm,nsd,nsd2,nsd3,Hill:1991at,Hill:1994hp,nsd4,littlehiggs,ed,rs1,rs2}.
These interactions predict new massive states,
generically referred to as
\zp bosons, that can decay into $\ttbar$ pairs.
Typical examples are the topcolor \zp described in
Refs.~\cite{nsd3,Hill:1991at,Hill:1994hp}, and the Randall--Sundrum Kaluza--Klein (KK) gluons
of Ref.~\cite{rs_gluon_1}.
Other models~\cite{bhkr,axigluon_ttbar_1,axigluon_ttbar_2,axigluon_ttbar_3}
have recently been proposed
to resolve the discrepancy in the forward-backward asymmetry
in $\ttbar$
production reported at the Tevatron
\cite{top_AFB1,top_AFB2,top_AFB3,top_AFB4,top_AFB5}.
Model-independent studies of the implications
of a large forward-backward
asymmetry 
suggest that a strong enhancement of the production cross
section for $\ttbar$ pairs would be expected at the Large Hadron Collider (LHC) for invariant
masses $m_{\ttbar} > 1$\TeVcc, if the observed discrepancy with the
predictions of the standard model (SM) is due to
new physics at some large mass scale~\cite{top_afb_implications1,top_afb_implications2}.
Searches for new physics in top-pair production have been performed
at the Tevatron~\cite{cdftt1, cdftt2, d0tt}, and
provide the most stringent lower limits on the mass ($\mzp$) of
narrow-width ($\Gamma_{\zp}$)
resonances, e.g. excluding a topcolor $\ttbar$ resonance with
$\Gamma_{\zp} / \mzp = 1.2\%$ for masses below $\approx 0.8$\TeVcc.

In this Letter, several models of resonant $\ttbar$ production
are considered, including a $\zp$
resonance with a narrow width of 1\% of the mass,
a $\zp$ resonance with a moderate width of 10\% of the mass, as well as
broader KK
gluon ($\mathrm{g'}$) states~\cite{rs_gluon_1}.
An enhancement over $\ttbar$ continuum
production at large $\ttbar$ invariant masses is also considered.

This study examines decays of produced $\ttbar$ pairs in the
all-hadronic channel, taking advantage of the large (46\%) branching
fraction of $\ttbar \rightarrow \wboson^+ \cPqb \wboson^- \cPaqb \rightarrow$ 6 quarks,
and focuses on final states with $m_{\ttbar} > 1$\TeVcc.
For lower masses, the background from quantum-chromodynamic (QCD)
production of non-top multijet events makes the
search prohibitively difficult in this channel.  At high $m_{\ttbar}$, using new techniques in
jet reconstruction to identify jet
substructure~\cite{catop_theory,catop_cms,jetpruning1,jetpruning2,boostedhiggs},
it is
possible to study highly boosted top quarks ($E/m_\mathrm{t}c^2 > 2$, where
$E$ and $m_\mathrm{t}$ are the energy and mass of the top quark).
The decay products of these
highly boosted top quarks are collimated, and are partially or fully
merged into single jets with several separate
subjets corresponding to the final-state quarks (one from the
bottom quark, and two light-flavor quarks from the $\wboson$ decay).
The data sample
corresponds to an integrated luminosity of ~\intlumi\ collected by
the Compact Muon Solenoid (CMS) experiment~\cite{:2008zzk} in
proton-proton
collisions at a center-of-mass energy of 7\TeV at the LHC.

In the following Letter, Sec.~\ref{sec:cms_detector}
describes the CMS detector and the event reconstruction.
Section~\ref{sec:analysis_overview} explains the strategy
for the analysis and the derivation of the efficiency and
misidentification probability of the substructure
tools that were used. Section~\ref{sec:systematics} gives the
systematic uncertainties in the analysis.
Section~\ref{sec:statistics_paper} describes the statistical methodology
used. Section~\ref{sec:conclusions} presents a summary of
the results.

\section{CMS detector, event samples, and preselection}

\ifnpas
\section{CMS Detector}
\fi

\label{sec:cms_detector}

The CMS detector is a general-purpose detector that uses a silicon
tracker, as well as
finely segmented lead-tungstate crystal electromagnetic (ECAL) and
brass/scintillator hadronic (HCAL) calorimeters.
These subdetectors have full azimuthal coverage
and are contained within the bore of a superconducting solenoid that
provides a 3.8\unit{T} axial magnetic field. The CMS detector uses a polar coordinate system with the
polar angle $\theta$ defined relative to the direction ($z$) of the
counterclockwise proton beam.
The pseudorapidity $\eta$ is defined
as \(\eta = -\ln\tan(\theta/2)\), which agrees with the
rapidity \(y = \frac{1}{2} \ln \frac {E + p_{z}c}{E - p_{z}c}\) for
objects of negligible mass, where $E$ is the energy and
$p_{z}$ is the longitudinal momentum of the particle.
Charged particles are reconstructed in the tracker for $|\eta| < 2.5$.
The surrounding ECAL and HCAL provide coverage
for photon,
electron, and jet reconstruction for $|\eta| < 3$. The CMS
detector also has extensive forward calorimetry that is not used in
this analysis.
Muons are measured in gas-ionization detectors embedded in the steel return yoke outside the solenoid.

\label{sec:samples}

Events were selected with an online trigger system, with decisions based
on the transverse momentum ($\pt$) of a single jet measured in
the calorimeters.
The instantaneous luminosity increased with time, hence
two thresholds were used for different running periods.
Most of the data were collected with a threshold of jet $\pt > 300$\GeVc,
and the rest with a threshold of 240\GeVc.
Offline, one jet 
is required to satisfy $\pt > 350$\GeVc.

There are several Monte Carlo (MC) simulated samples in the
analysis.
The continuum SM $\ttbar$ background is simulated with the
\textsc{MadEvent}/\textsc{MadGraph}~4.4.12~\cite{MadGraph} and
\PYTHIA~6.4.22~\cite{pythia} event
generators.
The \textsc{MadGraph} generator is also used to model generic high-mass
resonances decaying to SM top pairs.  In particular, a model is
implemented with a \zp
that has SM-like fermion couplings and mass between
1 and 3$\TeVcc$. However, in the MC generation of the \zp,
only decays to $\ttbar$ are simulated.
The width of the
resonance is set to 1\% and 10\% of \mzp, so as to check the predictions for a narrow and a
moderate resonance width, respectively. Here, the 10\% width is
comparable to the detector resolution.
The \PYTHIA~8.145 event generator~\cite{pythia8} is used to generate Randall--Sundrum KK gluons
with masses $m_\mathrm{g'} = 1,$ 1.5, 2, and 3\TeVcc, and widths of
$\approx 0.2 m_\mathrm{g'}$.
These Randall--Sundrum gluons have branching fractions to
$\ttbar$ pairs of 0.93, 0.92, 0.90, and 0.87, respectively.
\PYTHIA{}6 is also used to generate non-top multijet events for background
studies, cross-checks, and for calculating correction factors.
The {CTEQ6L} \cite{cteq} parton distribution functions (PDF) are
used in the simulation. The detector response is simulated using the
CMS detector simulation based on \GEANTfour~\cite{Geant4}.

Events are reconstructed using the particle-flow
algorithm~\cite{particleflow}, which identifies all reconstructed
observable particles (muons, electrons, photons, charged and neutral hadrons)
in an event by combining information from all
subdetectors.
Event selection begins with removal of beam background
by requiring that events with at least 10 tracks have at least
25\% of the tracks satisfying high-purity tracking requirements \cite{highpuritytracks}.
The events must have a well-reconstructed primary vertex, and only
charged particles identified as being consistent with the highest $\Sigma \pt^2$
interaction vertex are considered, reducing the
effect of multiple interactions per beam crossing (pile-up) by
$\approx 60$\%.

The selected particles, after removal of charged hadrons from pile-up
and isolated leptons, are clustered into
jets using the Cambridge-Aachen (CA) algorithm with a distance parameter of
$R = 0.8$ in $\eta$-$\phi$ space, where $\phi$ is the azimuthal
angle~\cite{CAaachen,CAcambridge}, as implemented
in the \textsc{FastJet} software package version 2.4.2~\cite{fastjet1,fastjet2}.
The CA algorithm sequentially merges into single objects, by four-vector addition, the pairs
of particle clusters that are closest in the distance measure
\(d_{ij} = \Delta R_{ij}^2 / R^2\), where 
\(\Delta R_{ij} = \sqrt{(\Delta \eta)^2 + (\Delta \phi)^2}\) 
and $R = 0.8$, until the minimum is less than or
equal to the so-called beam distance \(d_{iB}\), which equals unity
in the CA algorithm. More generally these distance measures are equal to
\(d_{ij} = \mathrm{min}({\pt}_i^{2n},{\pt}_j^{2n}) \Delta R_{ij}^2 / R^2\)
and \(d_{iB} = {\pt}_i^{2n}\). In the more common cases of the
$k_\mathrm{T}$ and anti-$k_\mathrm{T}$ algorithms~\cite{ktalg}, $n=1$ and $n=-1$,
respectively, however for the CA algorithm $n=0$, and hence
only angular information is used in the clustering.
When the beam distance for particle cluster $i$ is
smaller than all of the other $d_{ij}$, particle cluster $i$ is identified as a
jet and the clustering proceeds for the remaining particles in the
event.
Jet energy scale corrections are applied as documented in Ref.~\cite{jec_jinst}.
All jets are required to satisfy jet-quality
criteria~\cite{particleflow}, as well as $|y| < 2.5$. The rapidity
is used in this case because the jets
acquire a finite mass as part of the imposed jet-quality criteria.

\section{Analysis method}

\label{sec:analysis_overview}

The analysis is designed for cases in which the
$\ttbar$ system has sufficient energy for
the decay products of each top quark to be emitted into a single
hemisphere,
implying that
$E/m_{\mathrm t}c^2 > 2$. As a consequence, the top
quarks can be either partially merged when only the $\wboson$ decay
products are merged
into a single jet,
or fully merged when all top decay products are merged into a single
jet. Thus, this analysis
becomes inefficient for low masses, and upper
limits are not evaluated for $\zp$ masses below 1\TeVcc.

The largest background in this search is the non-top multijet
(NTMJ) background. This is highly suppressed by requirements on the
jet mass and substructure.
The remaining NTMJ background is
estimated by computing the probability for non-top jets
to pass the top-jet selections (misidentification probability) in
control regions of the data. These control regions are constructed
by inverting substructure selections
while keeping mass selections fixed. This mistagging probability
is then applied to
the signal region to estimate the contribution from the NTMJ background.

In this section, the jet topologies in the analysis are defined
in Sec.~\ref{sec:analysis_jettopologies}, the signal estimate
is described in Sec.~\ref{sec:signalestimate}, and the
background estimate is shown in Sec.~\ref{sec:backgroundestimate}.
Finally, the results of the event selection and the background
estimate are presented in Sec.~\ref{sec:results_of_event_selection}.

\subsection{Analysis of jet topologies}
\label{sec:analysis_jettopologies}

The events are classified into two categories,
depending on the number of final-state jets that
appear in each hemisphere. The 1+1 channel
comprises dijet events in which each jet corresponds to a fully merged top-quark candidate,
denoted as a Type-1 top-quark candidate.
The 1+2 channel comprises trijet events that fail the 1+1 criteria,
with a Type-1 top-quark candidate in one
hemisphere, and at least two jets in the other, one being a jet from a
$\cPqb$ quark
(although no identification algorithms are applied) and the
other a merged jet from a $\wboson$.
These two separate jets define
a Type-2 top-quark candidate in the 1+2 channel.
Further channels, such as 2+2, which would correspond to two Type-2
top quarks, are not considered in this analysis.
The 1+1 and 1+2 selections are now discussed in detail.

The 1+1 events are required to have at least two Type-1 top-quark
candidates, each reconstructed with $\pt > 350\GeVc$. Both
candidates are tagged by a top-tagging
algorithm~\cite{catop_theory,catop_cms}
to define merged top jets.
In the case of more than two top-tagged jets, the two top-tagged jets with the highest
$\pt$ are considered.
The top-tagging algorithm is based on the
decomposition of a jet into subjets, by reversing the
final steps of the CA
jet-clustering sequence.
In this decomposition, particles that have small $\pt$ or are at large angles
relative to the parent cluster are ignored. At least three
subjets are required in each jet. While the subjets of generic
jets tend to be
close together, and one of them often dominates the jet energy because of gluon
emission in the final state, the decay products of the top quark
share the jet energy more equally and emerge at wider angles. The mass
of the summed four-vector of the constituents of the hard jet must
be consistent with the mass of a top quark $m_{\mathrm{t}} \approx
175$\GeVcc ($140 < m_{\text{jet}} < 250$\GeVcc, where the values chosen are optimized through MC simulation).
Figure~\ref{figs:jetMass_DataMC} shows the expected jet mass
for the $\zp$ signal from MC as a dotted histogram, and the expected jet
mass for the NTMJ background from MC as a solid yellow histogram.
As expected, the $\zp$ signal has a peak at the top mass corresponding
to fully merged top jets, and
has a shoulder at the $\wboson$ mass corresponding to partially merged
top jets.
The minimum pairwise invariant mass of the three subjets
of highest $\pt$ is required to be
$>50$\GeVcc, because the combination with the
minimum pairwise mass often ($> 60\%$) consists of the jet
remnants from the $\wboson$ decay.

The 1+2 events are required to have exactly one hemisphere containing
a top-tagged Type-1 candidate with $\pt > 350\GeVc$. That is,
only events that fail the 1+1 criteria are considered for the
1+2 selection.
In the hemisphere opposite the top-tagged Type-1 candidate,
there must be at least two jets, one identified as a
$\wboson$-jet candidate, with $\pt > 200\GeVc$, and another jet
from a $\cPqb$ quark (although no identification algorithm is
used) with $\pt > 30\GeVc$. The
$\wboson$ jet is required to be tagged by a
$\wboson$-tagging algorithm, based on the jet pruning
technique~\cite{jetpruning1,jetpruning2}. The $\wboson$-tagging
algorithm requires two subjets, a total jet mass consistent with the mass of the
$\wboson$ boson $m_\wboson = 80.4$\GeVcc (\(60 < m_{\text{jet}} < 100\)\GeVcc), and
an acceptable ``mass-drop'' parameter $\mu$ of the final
subjets relative to the hard jet~\cite{boostedhiggs}. The mass-drop $\mu$ is defined as the ratio of the
mass of the more massive subjet $m_1$ to the mass of the complete jet
$m_\text{jet}$, and is required to be smaller than 40\%
(\(m_1/m_\text{jet} \equiv \mu < 0.4\)).
This selection helps to discriminate against generic jets, which
usually have larger $\mu$ values.
The $\wboson$-jet and $\cPqb$-jet candidates combine to
form the Type-2 top-quark candidate, whose mass must be consistent
with that of the top quark
(\(140 < m_{\text{jet}} < 250\)\GeVcc,
where the values chosen are optimized using MC simulation).
When there are more than two jets in the Type-2 hemisphere, the
$\cPqb$-quark candidate is taken as the one closest to the
$\wboson$-tagged jet in $\eta$-$\phi$ space.

The jet-pruning technique used to select $\wboson$ jets removes a portion of the jet,
which is not accounted by the ordinary jet energy scale corrections. The jet
corrections used in this analysis are derived from unpruned jets,
and the impact on pruned jets is therefore investigated using a
dijet MC sample. In particular, the $\pt$ of reconstructed pruned
jets are compared to the $\pt$ of matched generator-level particle jets,
that also underwent the pruning procedure, and
the difference of 2--3\%
observed in absolute response
suggests a systematic uncertainty in the jet energy from this source.
An uncorrelated
3\% uncertainty is therefore added to the uncertainties for
standard jet energy corrections. This uncertainty is applied for
both the top-tagging and jet-pruning algorithms, and is added in quadrature to the general jet
energy scale corrections of Ref.~\cite{jec_jinst}, which are
$\approx 2$--$4$\%, depending on $\pt$ and rapidity.

\subsection{Signal efficiency}
\label{sec:signalestimate}

\begin{figure}
\centering
\subfigure[]{\label{figs:jetMass_DataMC}\includegraphics[width=0.65\textwidth]{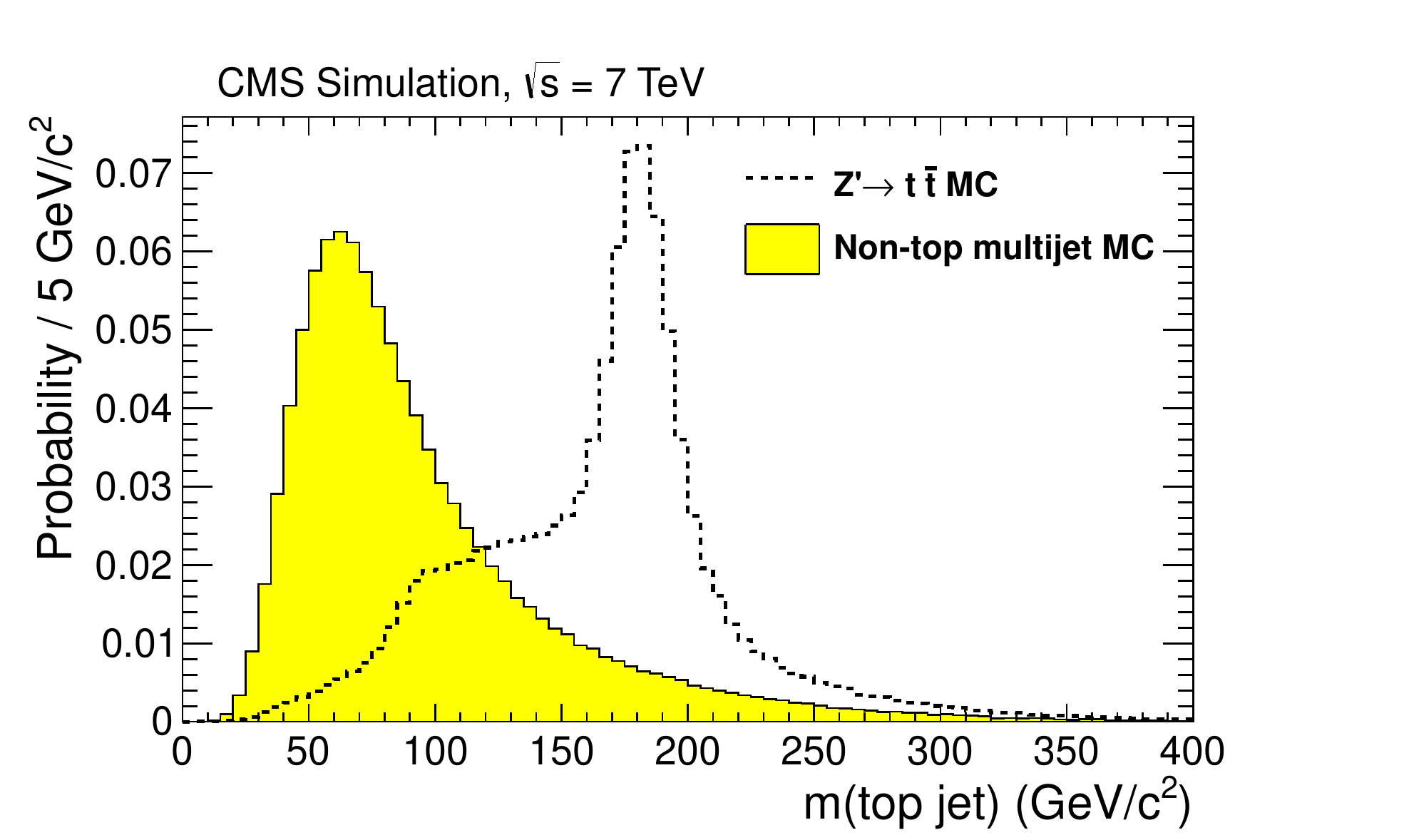}}\\
\subfigure[]{\label{figs:TOP_MISTAG_RATE}\includegraphics[width=0.65\textwidth]{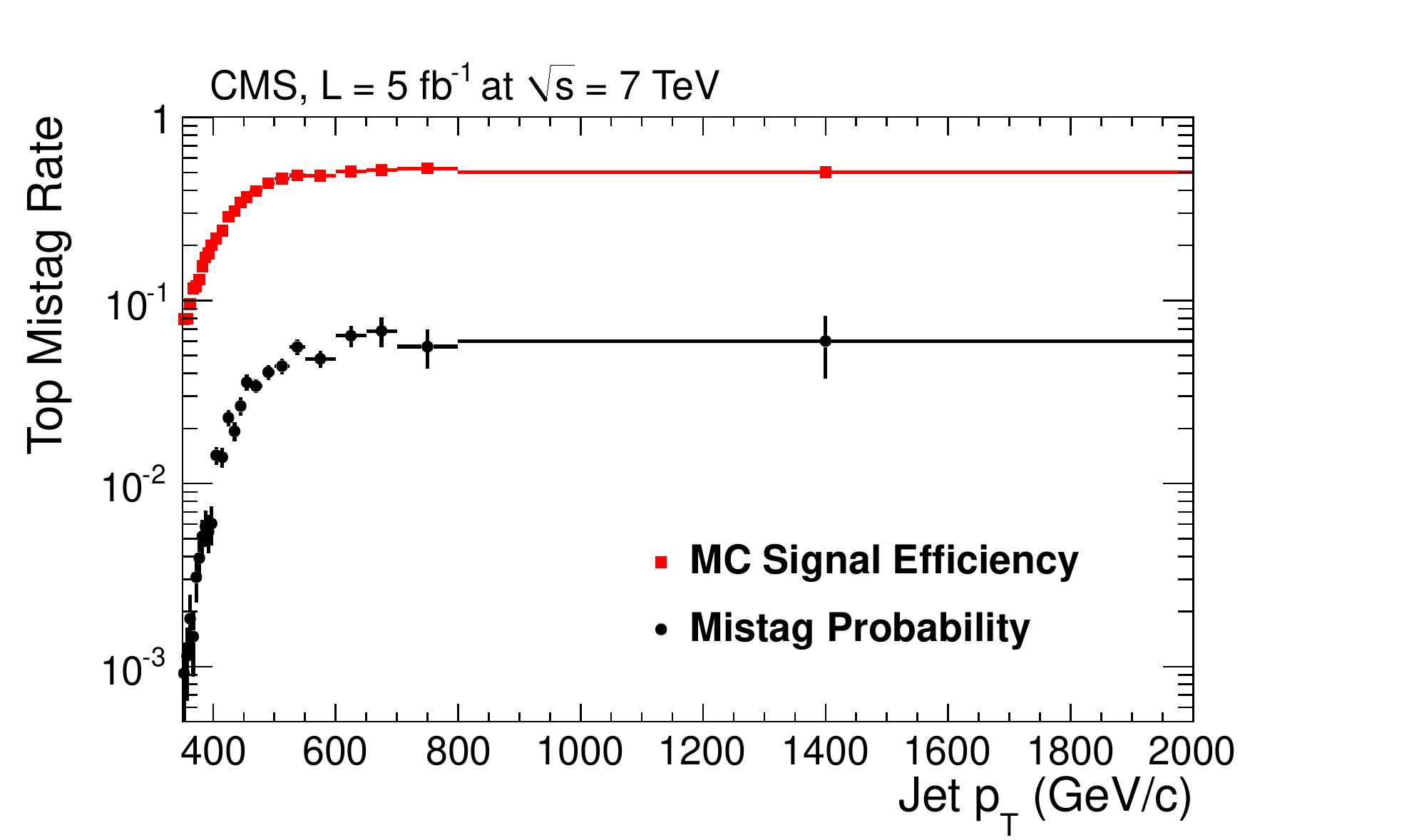}}
\caption{
(a) The simulated jet mass for NTMJ MC (light yellow histogram) and
$\zp$ MC (open histogram).
(b) The type-1 per-jet top-tagging efficiency for $\zp$
MC events is shown as red squares with error bars (see Sec.~\ref{sec:signalestimate}), and
the type-1 per-jet mistag probability for top tagging measured in data 
is shown as black circles with error bars (see Sec.~\ref{sec:backgroundestimate}), both as a function of jet $\pt$. }
\end{figure}

For both the $\zp$ signal and the (subdominant) $\ttbar$ background, the
efficiencies of the analysis selection are estimated from Monte Carlo
simulation, with scale factors
to account for measured differences with
respect to the data.
Figure~\ref{figs:TOP_MISTAG_RATE} shows the efficiency for tagging a
true top jet as a function of $\pt$.
Above $\approx 500\GeVc$, the efficiency plateaus between 50--60\%.

Three scale factors are applied to the efficiency.
The first scale factor is used to correct for the trigger in simulated
signal events.
Its value is equal to the trigger efficiency: it rises from $\approx 75\%$
(60\%) for 1+1 (1+2) events at  $m_{\ttbar} = 1.0$\TeVcc and
becomes fully efficient for $m_{\ttbar} > 1.5$\TeVcc.
The value of the trigger efficiency is estimated per jet
on a sample of simulated NTMJ events
passing the top-quark candidate selections,
it is then applied to simulated signal events.
The systematic uncertainty is assigned to be 50\% of the trigger
inefficiency from MC.
The difference between the measured trigger efficiency in data and MC
is roughly in this range,
but suffers from large statistical uncertainties.
The second
scale factor is used to correct for
any differences in jet-energy scale for the subjets and for the full jets.
This is referred to as the subjet jet-energy scale factor.
The third scale factor is used to correct for the impact of any
differences between data and MC in efficiencies for finding jets with
substructure. This is referred to as the subjet selection-efficiency
scale factor.
The derivation of the second and third scale factors is now discussed in detail.

These second two scale factors are both determined in
a control sample comprising events with a single muon (referred to as the muon
control sample), usually from the decay of
$\mathrm{t} \rightarrow \wboson \mathrm{b}$, with
$\wboson \rightarrow \mu \nu$, and at least two jets.
The event selection is nearly identical to that in
Ref.~\cite{TOP-10-003}, except for larger $\pt$ requirements
for the jets. The leading jet is required to satisfy $\pt > 200$
\GeVc, and the sub-leading jet is required to satisfy $\pt > 30$ \GeVc. 
Due to a small number of
fully-merged (Type-1) top jets in this sample, it is not possible to construct a 
sufficiently large number of true Type-1 top jets. 
Instead, the characterization of jets with substructure is studied
with moderately-boosted $\ttbar$ events
where there is a large fraction of partially-merged (Type-2) top-quark
candidates and the $\wboson$-jets within
them. 

The events in the muon control sample used to study $\wboson$-tagging are dominated by $\ttbar$ decays,
and the leading-jet $\pt$ requirement ($> 200$ \GeVc) 
favors the topology in which the two
top quarks are produced back-to-back, thereby facilitating jet
merging on the side of the ``hadronic'' top quark, containing jets from $\mathrm{t} \rightarrow
\mathrm{W} \mathrm{b} \rightarrow \mathrm{q}\mathrm{\overline{q}{}'}\mathrm{b}$.
In the other hemisphere, these events contain one
isolated muon, consistent with originating from the primary collision
vertex, with $\pt>45\GeVc$ and $|\eta|<2.1$. Events are rejected if
they contain other isolated
electrons or muons within $|\eta|<2.1$ with $\pt>15\GeVc$ or $\pt>10\GeVc$,
respectively. The remaining events
are then required to have $\ge 2$ jets with $\pt >
30\GeVc$, and at least one jet with $\pt > 200\GeVc$.
Unlike the all-hadronic channel discussed
in this Letter, in the muon control sample,
there are two well-separated jets originating from $\cPqb$ quarks,
so
in order to enhance the $\ttbar$ fraction of this control region,
the events are required to contain at least one jet tagged with
a secondary-vertex b-tagging algorithm~\cite{CMS-PAS-BTV-11-001}, also
used in Ref.~\cite{TOP-10-003}.
The b-tagging algorithm combines at least two tracks
into at least one secondary vertex, and forms a discriminating
variable based on the three-dimensional decay length of the vertex.

\begin{figure}
\centering
\subfigure[]{\label{figs:wMass_semileptonic}\includegraphics[width=0.65\textwidth]{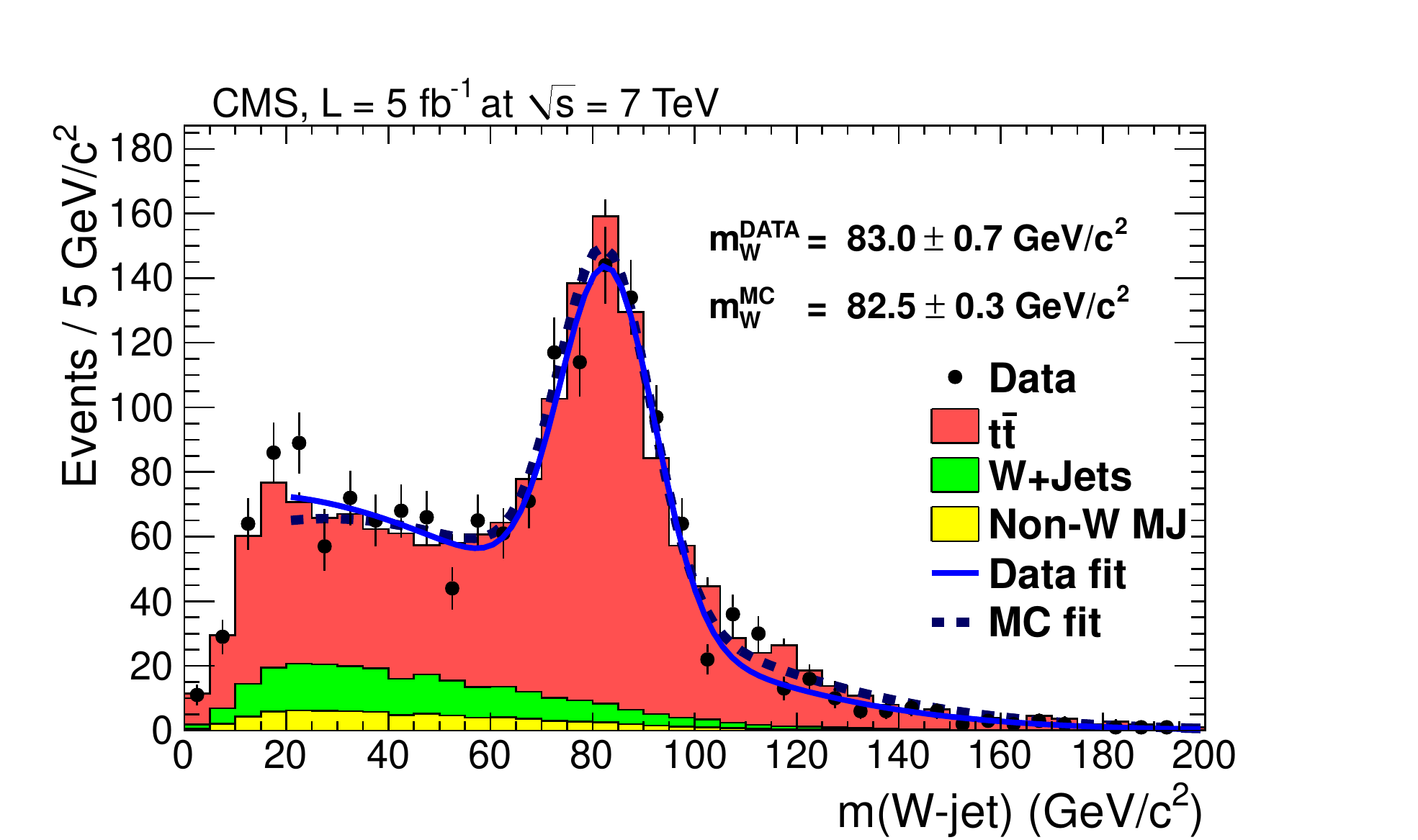}} \\
\subfigure[]{\label{figs:topMass_semileptonic}\includegraphics[width=0.65\textwidth]{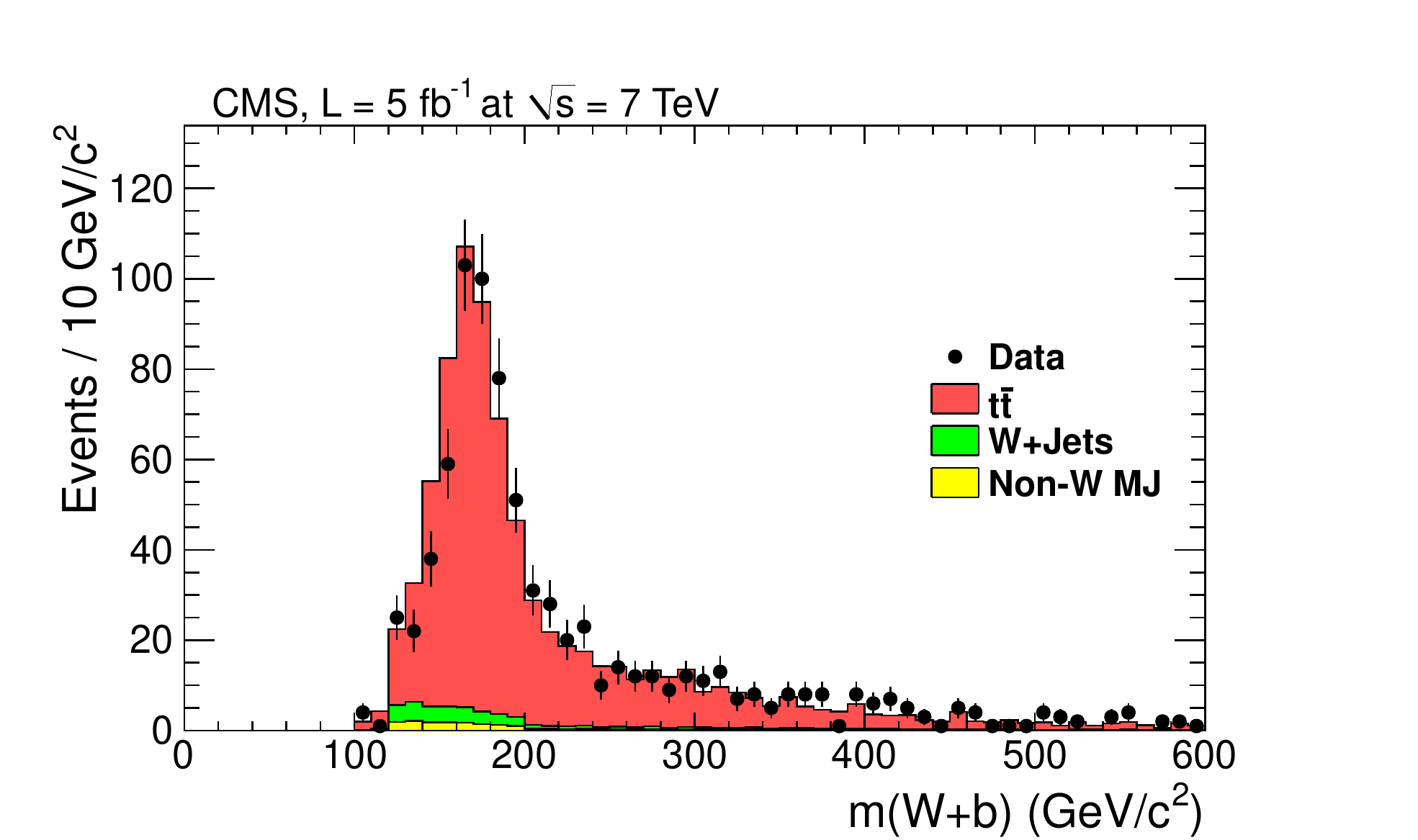}}
\caption{ (a) The mass of the highest-mass jet ($\wboson$-jet),
  and (b) the mass of the Type-2 top candidate ($\wboson+\cPqb$),
  in the hadronic hemisphere of moderately-boosted events in the muon
  control sample. The data are shown as
  points with error bars, the $\ttbar$ Monte Carlo events in dark red, the $\wboson+$jets Monte
  Carlo events in lighter green, and non-$\wboson$ multijet
  (non-$\wboson$ MJ) backgrounds are
  shown in light yellow (see Ref.~\cite{TOP-10-003} for details of non-$\wboson$ MJ
  distribution derivation).
  The jet mass is fitted to a sum of two Gaussians in
  both data (solid line) and MC (dashed line), the latter of which lies directly
  behind the solid line for most of the region.}
\end{figure}

The subjet jet-energy scale factor is estimated by extracting a
$\wboson$ mass peak in the muon control sample, and comparing the
peaks in data and MC.
The
mass distribution of the jet of largest mass in the hadronic hemisphere is
shown in Fig.~\ref{figs:wMass_semileptonic}. 

In this figure, the MC $\ttbar$ contribution is normalized to the approximate
next-to-next-to-leading-order (NNLO) cross section for inclusive
$\ttbar$ production of 
163\unit{pb}~\cite{Aliev:2010zk,Langenfeld:2009wd,Kidonakis:2010dk}.
The non-$\wboson$ mulitjet component is based on sidebands in data
that have the muon isolation criterion reversed. The contributing
spectrum is normalized through a fit to the missing transverse
energy.  
The stringent criteria of this analysis provide very few
$\wboson$+jets events that pass the required selections. This
distribution is therefore taken to be the same as that of the generic
non-$\wboson$ multijet background. This is acceptable because the mass
structure within the candidate top-quark jets are very similar in
these two samples, and the sideband data has many more events that
pass the selection criteria.
The $\wboson$+jets contribution is normalized to
the inclusive
$\wboson$ production cross section of 
$\sigma_{\wboson\to \ell+\nu} = 31.3 \pm 1.6~$nb computed at NNLO
with {\sc FEWZ}~\cite{fewz}.
A fit of the sum of
two Gaussian functional forms to data is given by the solid line, and
a similar fit to the simulated events is shown as a dashed line.
The centers
of the main Gaussian distributions in data and MC are $\wmasssemilepdata$ and
$\wmasssemilepmc$, respectively. The subjet jet-energy scale
factor for $\wboson$ jets is determined by taking the ratio of these two
values, and equals $\subjetjes$, including statistical uncertainty only.

The subjet selection-efficiency scale factor is
estimated by comparing the observed selection efficiency in the muon control sample in data
and MC.
The ratio of the number of events in the $\wboson$ mass window
($60 < m_{\mathrm{jet}} < 100$ \GeVcc) in
Fig.~\ref{figs:wMass_semileptonic}, after $\wboson$
tagging, to the number of events in the muon control sample, defines the
$\wboson$ selection efficiency within the mass window.
For data and MC the values are
$\mcuteffsemilepdata$ and $\mcuteffsemilepmc$, respectively. Similarly, the mass-drop
selection is checked in data and Monte Carlo, following the $\wboson$-mass window
selection, and a similar efficiency is extracted, with the observed values
being $\mucuteffsemilepdata$ and $\mucuteffsemilepmc$.
Combining efficiencies of the mass-drop and mass selections, the subjet
selection-efficiency scale factor, to be applied to the MC to obtain the same
efficiency as in data, is determined to be $\scalefactor$.
The same scale factor and uncertainty are assumed for Type-1 jets,
which is consistent with results from the statistics-limited control sample of muon
events that contain Type-1 jets. 
As two top-quark tags are required in each event, the correction for
both 1+1 and 1+2 events is the square of the single-tag scale factor, yielding
$\scalefactorsqrd$.

The Type-1 jet selection cannot be checked at the same level of
precision as the $\wboson$-jet selection because of
the small number of Type-1 jets in the muon control sample.
However, the Type-2 top-quark candidate selection can be tested in the
muon control sample, as shown in Fig.~\ref{figs:topMass_semileptonic}.
The same procedure is used to construct these Type-2 top-quark candidates as in the
1+2 selection, including the $\wboson$-mass and mass-drop selections.
Within the statistical uncertainty,
good agreement is observed in the data and simulation. 
Since the selection of Type-2 top-quark candidates considers a boosted
three-body decay as well as a $\wboson$ tag, the agreement between
the characteristics of candidates in data and in MC provides further
confidence for the assumption that the scale factor for the efficiency
of $\wboson$-tagging is appropriate for three-body decays such as the
Type-1 top-quark system.

To check the dependability of our assumption, namely that the
data-to-MC scale factors are the same for the muon control sample as
for the $\zp$ signal, the scale factor is measured in a control sample
using more stringent kinematic requirements to select a part of phase
space similar to that of a $\zp$ signal with $m \sim 1\TeVcc$. The
$\cPqb$-tagging requirement is dropped in order to collect more
$\ttbar$ events. Also, to capture the kinematics of the background,
instead of using the distribution for $\wboson$+jets from the
sidebands in data without isolated muon candidates, as is done in the fit to
the $\wboson$ mass, the distribution for $\wboson$+jets is taken
instead from the $\wboson$+jets MC. 

\begin{figure}
\centering
\subfigure[]{\label{figs:semiLepMassDrawPt_ptTopCand_200}\includegraphics[width=0.65\textwidth]{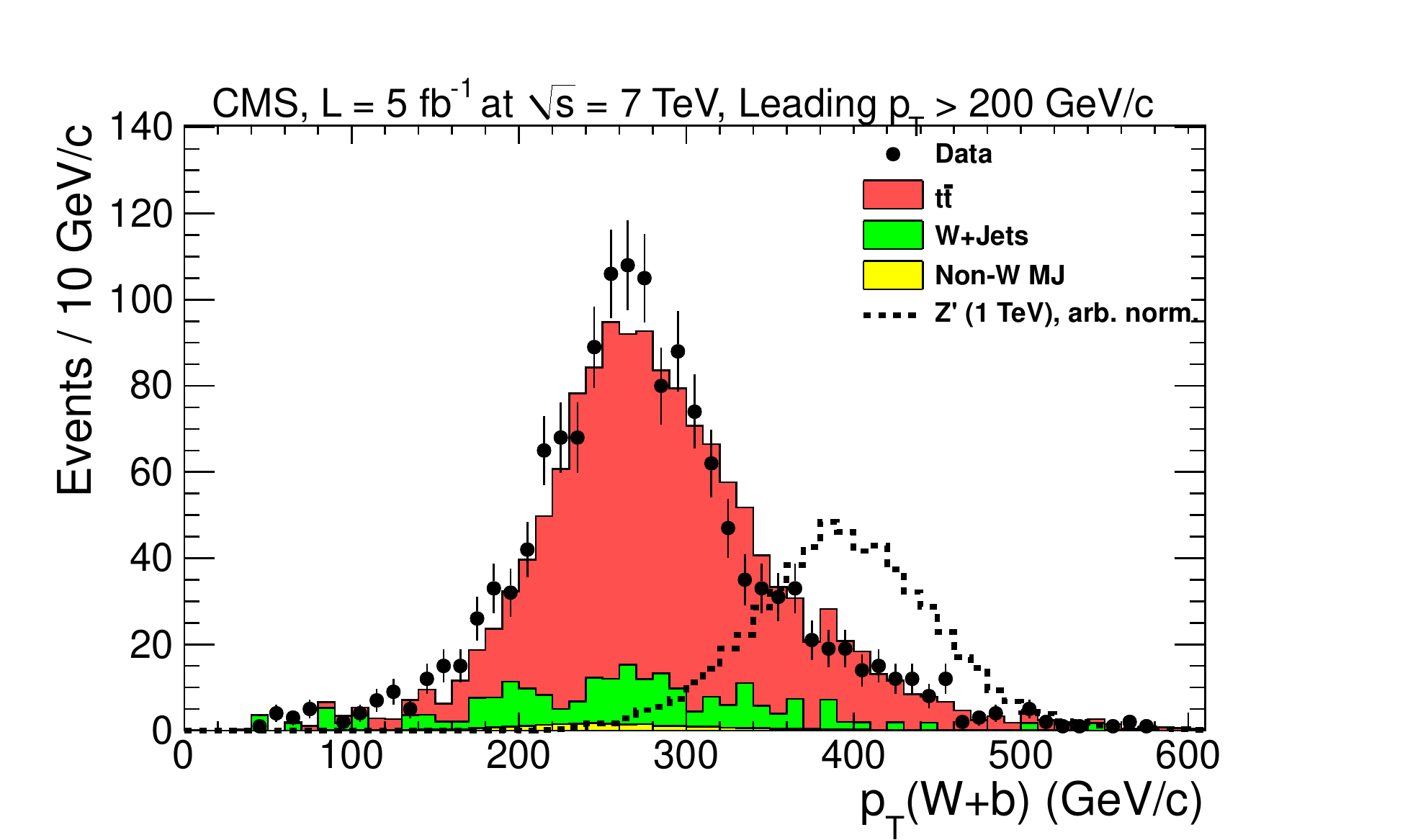}} \\
\subfigure[]{\label{figs:semiLepMassDrawPt_ptWFromTopCand_200}\includegraphics[width=0.65\textwidth]{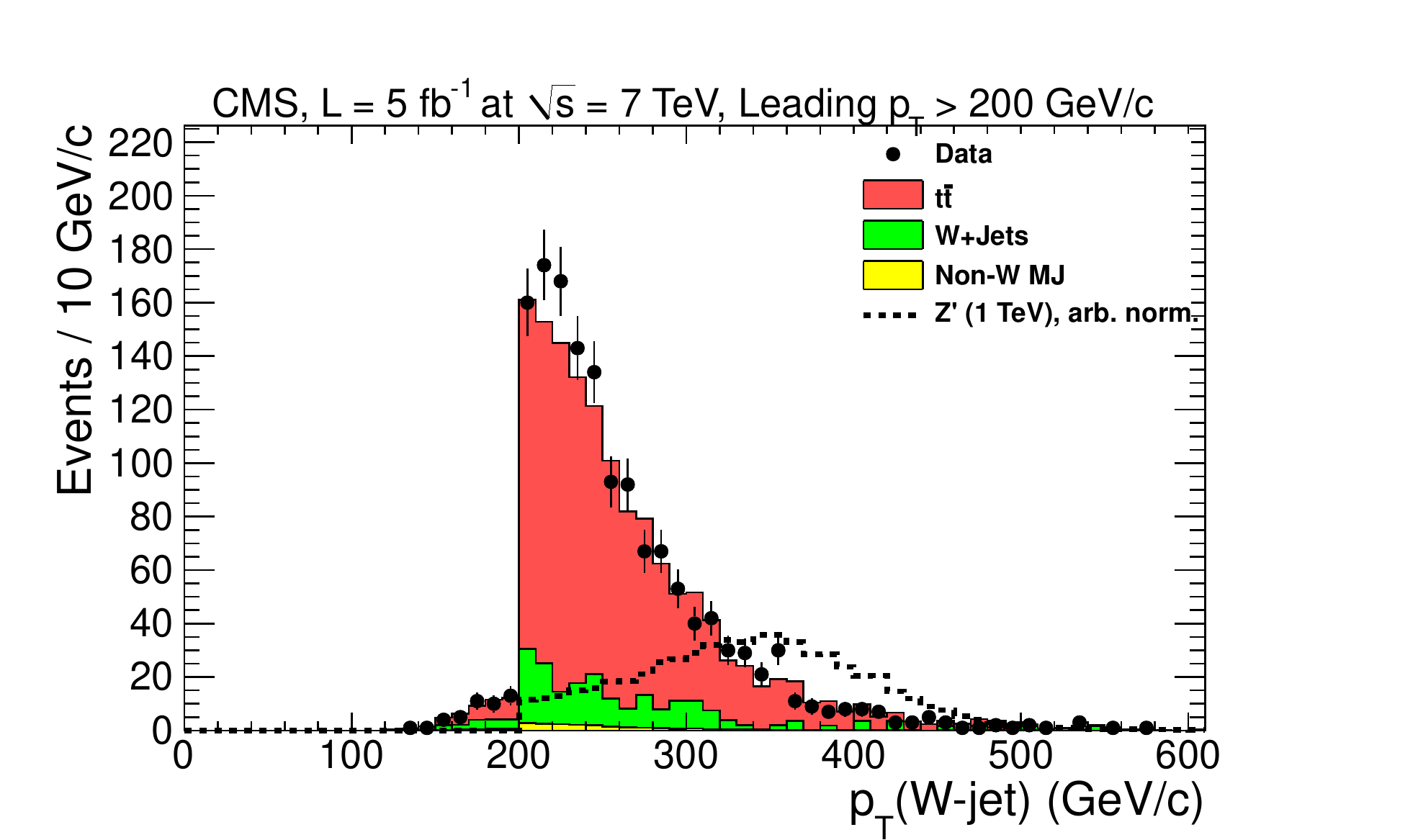}} 
\caption{ (a) $\pt$ of the Type-2 top-quark candidate in the muon
  control sample. The color scheme is the same as in
  Figs.~\ref{figs:wMass_semileptonic} and
  ~\ref{figs:topMass_semileptonic}. (b) $\pt$ of the
  $\wboson$-candidate from within the Type-2 top-quark candidate,
  after a selection on the jet mass of the highest-mass jet in the
  muon control sample. Overlaid on both (a) and (b) are the
  corresponding distributions from a $\zp$ MC signal with $m=1$ \TeVcc
  (with arbitrary normalization for visualization) to compare
  kinematics in the muon control region to the signal region.}
\end{figure}

\begin{figure}
\centering
\subfigure[]{\label{figs:highType2TopPt400_semiLepMass_mWCand}\includegraphics[width=0.65\textwidth]{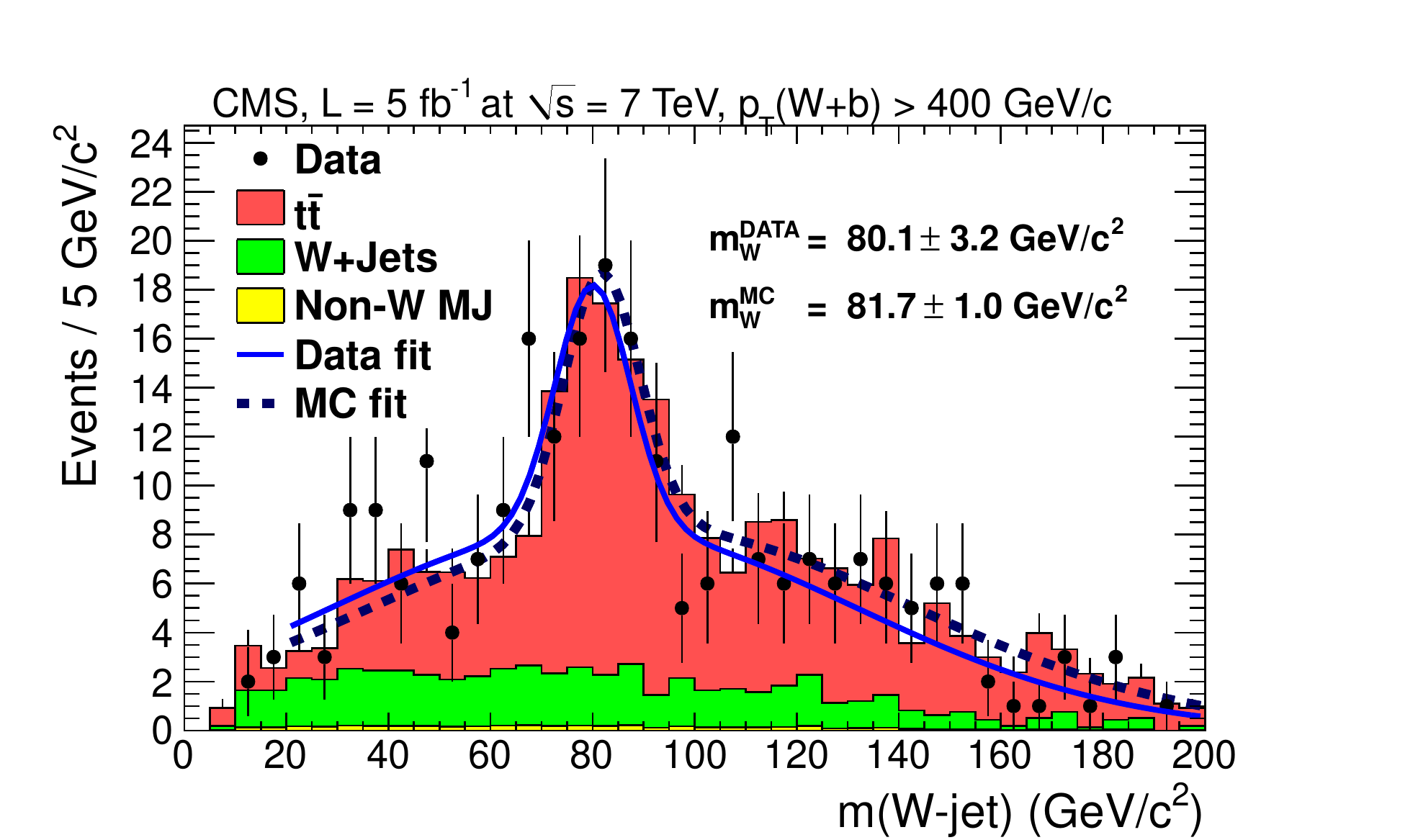}} \\
\subfigure[]{\label{figs:highType2TopPt400_semiLepMass_mTopCand}\includegraphics[width=0.65\textwidth]{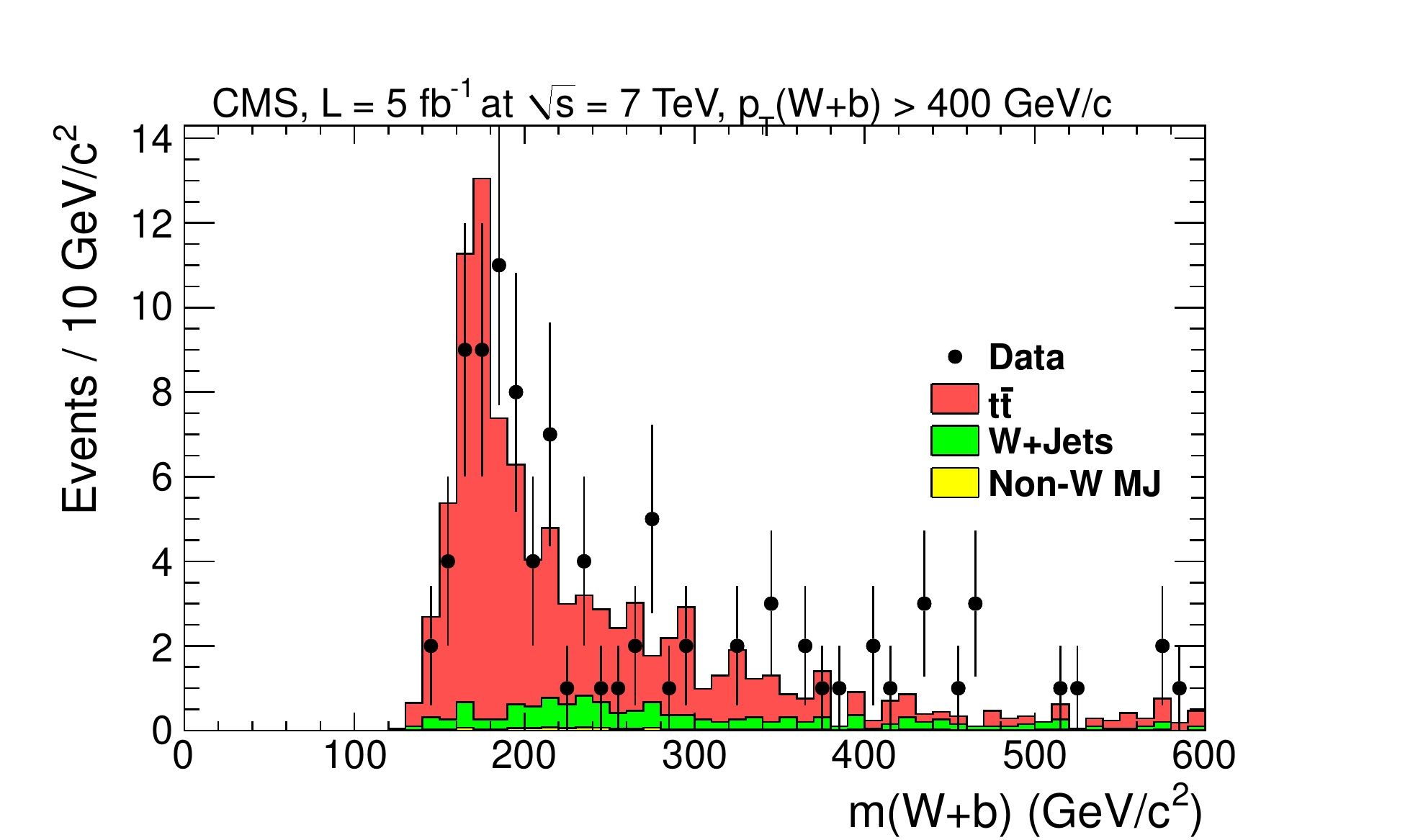}}
\caption{ (a) The mass of the highest-mass jet ($\wboson$-jet),
  and (b) the mass of the Type-2 top candidate ($\wboson+\cPqb$),
  in the hadronic hemisphere of moderately-boosted events in the muon
  control sample. The Figure corresponds to
  Figs.~\ref{figs:wMass_semileptonic}
  and~\ref{figs:topMass_semileptonic}, except there is an additional
  requirement on the Type-2 top candidate $\pt$ to be similar to the
  signal region. Figure~\ref{figs:semiLepMassDrawPt_ptTopCand_200}
  shows the distribution of the Type-2 top candidate $\pt$.}
\end{figure}

In all samples with large $\pt$ thresholds, the data-to-MC scale
factor is found to be consistent with the measured value of
$\scalefactor$. The selection that provides sideband regions most
similar in kinematics to that of the signal region is a requirement
that the Type-2 top candidate satisfy $\pt > 400$ \GeVc. 
Fig.~\ref{figs:semiLepMassDrawPt_ptTopCand_200} shows the $\pt$
distribution for the Type-2 candidates in the muon control sample, and
Fig.~\ref{figs:semiLepMassDrawPt_ptWFromTopCand_200} shows the $\pt$ of
the $\wboson$-jet within the Type-2 top-quark candidate, as defined by the jet
of largest mass in the event. Arbitrarily normalized distributions for
a $\zp$ signal with $m=1$ \TeVcc are overlaid for comparison. For
completeness, 
Figs.~\ref{figs:highType2TopPt400_semiLepMass_mWCand}
and~\ref{figs:highType2TopPt400_semiLepMass_mTopCand} show plots
identical to
Figs.~\ref{figs:wMass_semileptonic}
and~\ref{figs:topMass_semileptonic}, but with selections that require
Type-2 top-quark candidates with $\pt > 400$ \GeVc. The scale factor
extracted from this higher-$\pt$ subsample is 0.99
$\pm$ 0.11, which is consistent with the quoted data-to-MC scale factor of $\scalefactor$.

\subsection{Background estimate}
\label{sec:backgroundestimate}

Since this analysis focuses on signatures with high-$\pt$ jets,
the main backgrounds expected are from SM non-top multijet production and
$\ttbar$ production. The background from NTMJ
production is estimated from
sidebands in the data as described below. For the $\zp$ masses
considered in this analysis, the irreducible SM $\ttbar$ component is
significantly smaller than the NTMJ background contribution, and is
therefore
estimated from MC simulation using the same correction factors as
found for the $\zp$ MC described in Sec.~\ref{sec:signalestimate}.
It is normalized to the approximate
NNLO cross section described in Sec.~\ref{sec:signalestimate}.

In both 1+1 and 1+2 channels, estimates of the dominant NTMJ
background are obtained from data as follows.
First,
the probability is estimated for mistaking a non-top jet as a top-quark jet
through the top-tagging algorithm. This procedure defines the mistag probability
($P_\mathrm{m}$).
Higher momentum jets have a larger
probability to radiate gluons, and as the jet $\pt$
increases, they are more likely to have top-like
substructure and thereby satisfy a top tag~\cite{Ellis:2007ib}. 
The mistag probability is therefore obtained as
a function of total jet $\pt$, 
using the following procedure.  
Events with $\ge 3$ jets are selected for the 1+2
topology, with the three leading
jets in the event required to pass $\pt$
thresholds of 350, 200, and 30\GeVc, respectively, without
any requirements placed on the jet mass of the Type-1 candidate. The mass of the
$\wboson$-boson candidate within the Type-2 candidate is required to
fall within the $\wboson$-boson mass window, and the invariant mass of the Type-2 candidate
is required to fall within the top-quark mass window. However, the
mass-drop requirement is inverted ($\mu > 0.4$) to define a
signal-depleted sideband. The small contribution
from the SM $\ttbar$ continuum is subtracted using MC expectation,
and the mistag probability $P_\mathrm{m}(\pt)$
is defined by the fraction of Type-1
candidates that are top-tagged, as a function of their $\pt$. The resulting
mistag probability appears in Fig.~\ref{figs:TOP_MISTAG_RATE}.

\label{sec:mistag_rate}

Next, the 1+2 and 1+1 samples are defined using a loose selection:
(i) in trijet events, two jets in one hemisphere are required to pass the Type-2
selection and (ii) in dijet events, one randomly-chosen jet is required
to pass the Type-1 selection.  In both cases, the other
high-$\pt$ jet in the event (the probe jet) is not required to be
top-tagged.
These samples are dominated
by NTMJ events. For each event in the loose selection ($i$),
the probability that the event would pass the
full selection in the signal region ($P_{\mathrm{NTMJ}}^i$)
equals the mistag probability ($P_{\mathrm m}(\pt)$), evaluated
at the $\pt$ of the probe jet in event $i$ ($\pt^i$),

\begin{equation}
\label{eq:w_ntmj}
P_{\mathrm{NTMJ}}^i = P_{\mathrm m}(\pt^i).
\end{equation}

The total number of NTMJ events
($N_{\mathrm{NTMJ}}$) is then equal to the
sum of the weights from Eq.~\ref{eq:w_ntmj}.

\begin{equation}
\label{eq:ntmj}
N_{\mathrm{NTMJ}} = \sum_{i=1}^{N_\text{loose}} P_{\mathrm{NTMJ}}^i = \sum_{i=1}^{N_\mathrm{loose}} P_\mathrm {m}(\pt^i),
\end{equation}
where $N_\text{loose}$ is the number of events passing the loose
selection and the other quantities are defined above.

The ensemble of jets in the loose pretagged region have, on average, a lower jet mass
than the jets in the signal region. Consequently, the $m_{\ttbar}$
spectrum in this sideband is kinematically biased.
To emulate the event
kinematics of the signal region, the jet mass of the probe jet is
ignored, and instead it is set to a value randomly drawn from the
distribution of jet masses of probe jets from NTMJ MC events
in the range 140 to 250\GeVcc. 

This procedure is cross-checked on a NTMJ
MC sample to ensure that the methodology achieves closure.
In this cross-check, half of the events in the MC are used to derive
a mistag probability using the above procedure, and then used to
predict the expected number of tags for the remaining events, which is compared
to the observed number of tags in these events. The
observed and expected number of tags
agree within statistical uncertainties.

Possible biases in the calibration procedure from the presence of a
new $\zp$ have also been investigated in the analysis. For instance, a
$\zp$ signal with $m_{\zp} = 3$ \TeVcc and a width of $30$ \GeVcc
contributes less than 1\% to the events defined through the loose
selection criteria as well as to the sideband regions used to
determine the probability of mistagging jets. 

The uncertainty on this procedure
is taken as half the difference between the $m_{\ttbar}$ distributions
obtained using the modified and unmodified probe-jet masses.
Two choices of alternative prior distributions for the probe-jet mass
were investigated, the MC-based prior described above, and a flat
prior. The systematic uncertainty estimated with the current method is
slightly more conservative.

Table~\ref{table:mistagRateDerivation} provides an estimate for
the mistagged NTMJ background for two
$\ttbar$ mass windows: 0.9--1.1\TeVcc and 1.3--2.4\TeVcc. The first row
corresponds to the number of events observed in the loose selection
($N_{\mathrm{loose}}$ from Eq.~\ref{eq:ntmj})
to which the mistag probability is applied.
The second row corresponds to the number of expected events from
the mistagged NTMJ background in the signal region ($N_{\mathrm{NTMJ}}$ in Eq.~\ref{eq:ntmj}).
As can be observed in
Table~\ref{table:mistagRateDerivation}, the primary uncertainty on the
NTMJ background is from the systematic uncertainty assigned
to the procedure for modifying probe-jet masses.

\begin{table}
\begin{center}
\caption{The number of events observed in the loose selection
($N_{\mathrm{loose}}$), which is used as input to compute
the number of events predicted in the signal region for
the mistagged NTMJ background ($N_{\mathrm{NTMJ}}$).
Both appear in Eq.~\ref{eq:ntmj} and are described
in detail in Sec.~\ref{sec:backgroundestimate}.
Figure~\ref{figs:TOP_MISTAG_RATE} shows the value $P_{\mathrm m}(\pt)$ used in
Eq.~\ref{eq:ntmj}.
For $N_{\mathrm{NTMJ}}$,
the first and second uncertainties are statistical and systematic,
respectively.}
\label{table:mistagRateDerivation}
\begin{tabular}{ |l|cc|cc| }
\hline
          & \multicolumn{2}{c|}{$m_{\ttbar} = $0.9--1.1\TeVcc}  & \multicolumn{2}{c|}{$m_{\ttbar} = $1.3--2.4\TeVcc} \\
          & 1+1 & 1+2 & 1+1 & 1+2 \\
\hline
$N_{\mathrm{loose}}$ &
22015 &
70545 &
18401 &
30253 \\
$N_{\mathrm{NTMJ}}$ &
 443 $\pm$ 4 $\pm$ 22  &
1239 $\pm$ 6 $\pm$ 31 &
 741 $\pm$ 6 $\pm$ 30  &
 817 $\pm$ 6 $\pm$ 36  \\
\hline
\end{tabular}
\end{center}

\end{table}

\subsection{Results of event selection}
\label{sec:results_of_event_selection}

\begin{figure}
\centering
\subfigure[]{\label{figs:ttMassType11LOG}\includegraphics[width=0.65\textwidth]{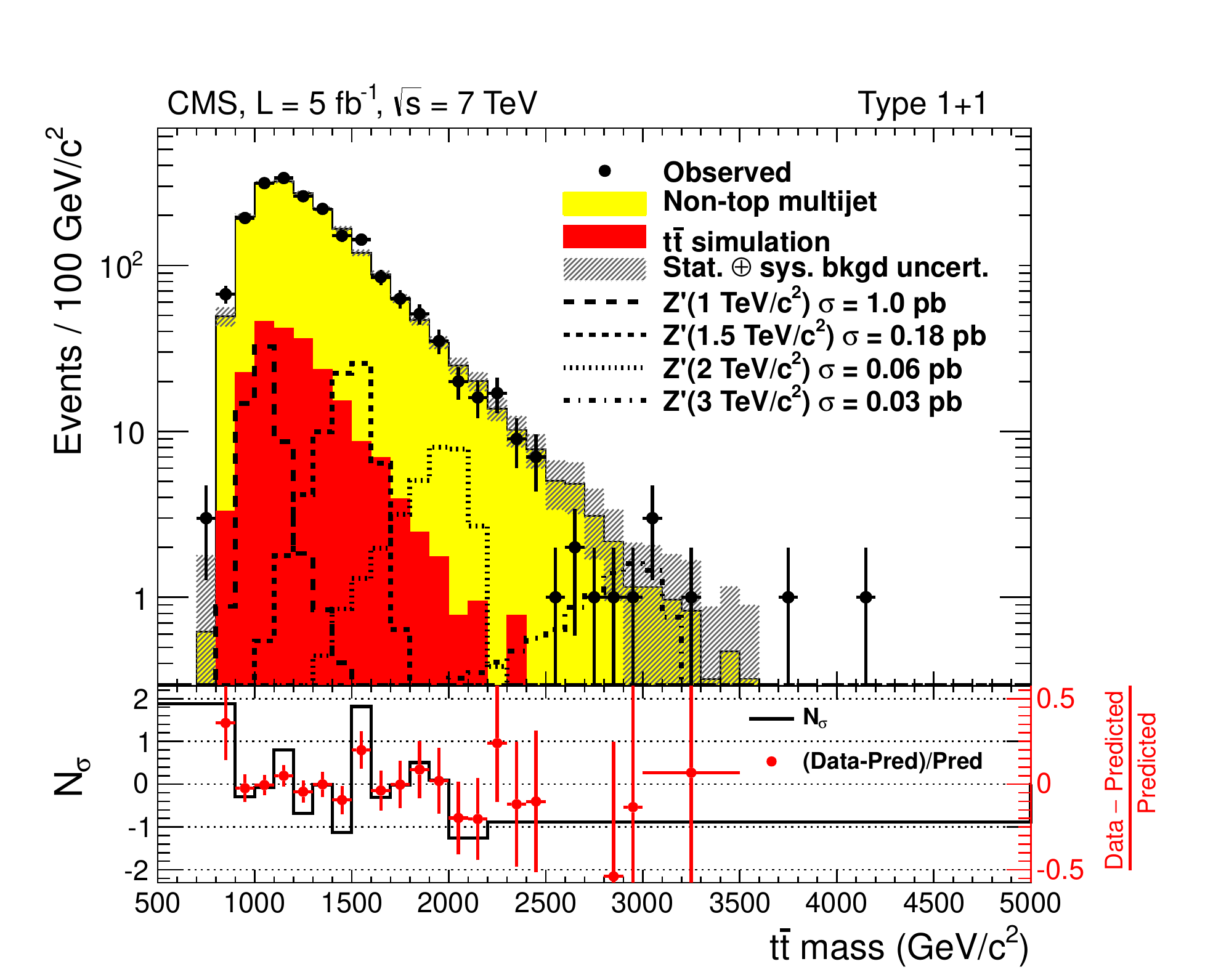}} \\
\subfigure[]{\label{figs:ttMassType12LOG}\includegraphics[width=0.65\textwidth]{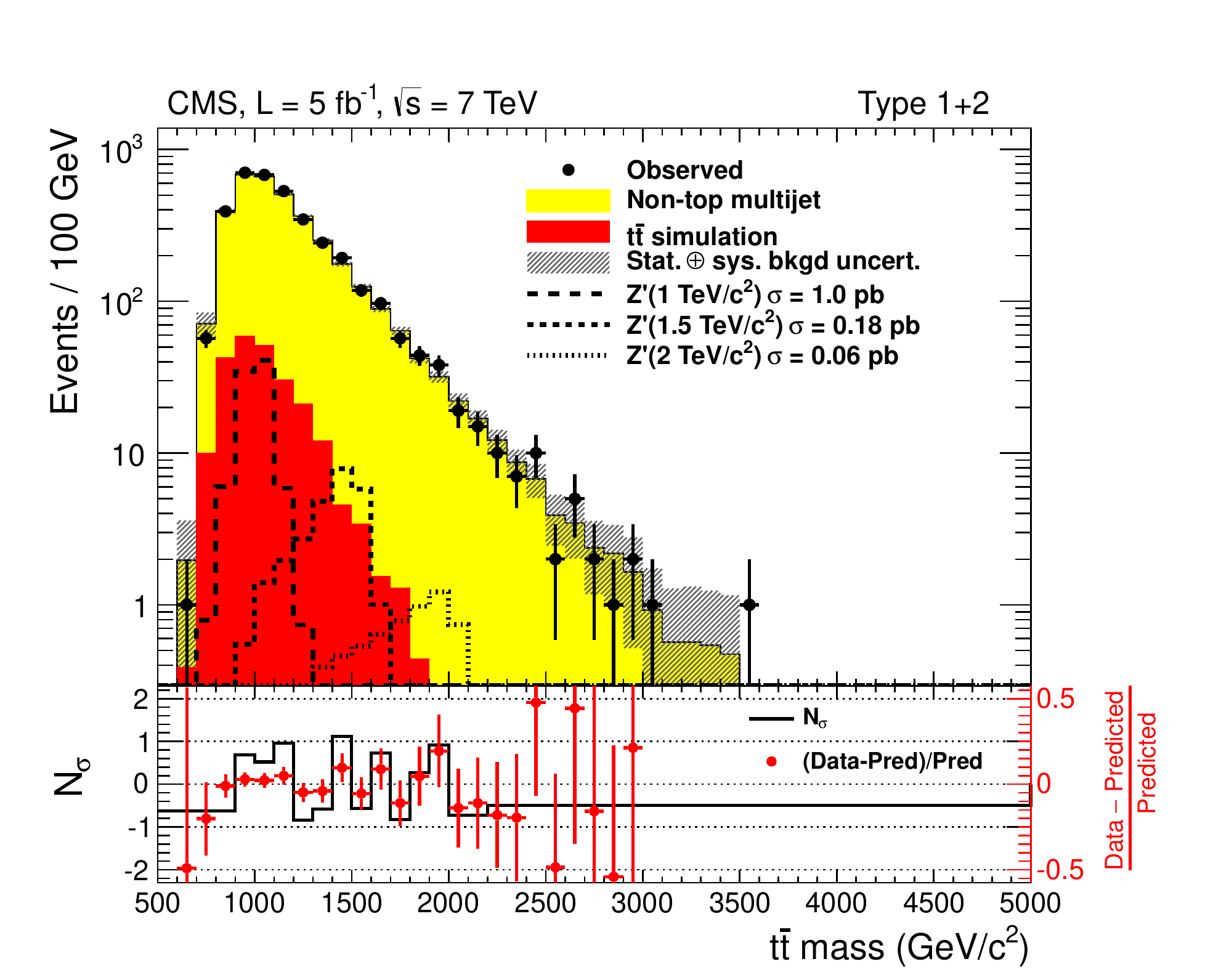}}
\caption{Results for (a) 1+1 and (b) 1+2 event selections and
  background estimates. The yellow (light) histograms are the non-top
  multijet (NTMJ) estimates from data,
  as described in the text, and the red (dark)
  histograms are the MC
  estimates from SM $\ttbar$ production.
  The black points are the data.
  The hatched gray boxes
  combine the statistical and systematic uncertainties on the total
  background. For comparison, expectations for some
  $\zp$ hypotheses are shown for the assumption of 1\% resonance width,
  with cross sections taken from
  the expected limits discussed in Sec.~\ref{sec:resonant}.
  Also shown are the ratio of the fractional difference between the data and the
  prediction, shown in red circles, with the $y$-axis on the right of the plot,
  and the number of standard deviations ($N_{\sigma}$) of the observation from the 
  prediction, shown as a black histogram with the $y$-axis on the left
  of the plot, where the binning is adjusted to have at least 20
  events in each bin.
  \label{fig:ttMassTwoTypes}
}

\end{figure}

Observed $m_{\ttbar}$ distributions for 1+1 and 1+2 events in data
are compared to the
expected backgrounds in Fig.~\ref{fig:ttMassTwoTypes}.
The NTMJ background expectation determined from data is
given by the yellow (light) filled histograms.
The SM $\ttbar$ estimate is shown as red (dark) filled
histograms, and the data are shown as solid black points.  The hatched
gray regions indicate the total uncertainty on the backgrounds.
Figure~\ref{fig:ttMassTwoTypes} also shows for comparison
the signal expectation from MC for several
hypothetical $\zp$ signals with masses from 1 to 3\TeVcc
with a width of 1\%,
in the 1+1 and 1+2 samples,
but with cross sections taken from
the expected limits discussed in Sec.~\ref{sec:resonant}.

From Fig.~\ref{fig:ttMassTwoTypes}, it is clear that the dominant
background in this analysis is from NTMJ events rather than from SM
$\ttbar$ production. The implementation of $\cPqb$-quark selections
has not as yet been introduced to improve the sensitivity of this
search, and must await an improvement in performance of
$\cPqb$-tagging in merged top jets. 

To demonstrate the components of the background estimate,
Table~\ref{table:countingExptCrossCheck} lists the number
of events expected from background sources in the 1+1 and
1+2 channels, along with the
observed number of events, for two $\ttbar$ mass windows: from
0.9--1.1\TeVcc, and 1.3--2.4\TeVcc.
The systematic uncertainties on these values are now summarized
in Sec.~\ref{sec:systematics}.

\begin{table}
\begin{center}
\caption{Expected and observed number of events in two different
  $\ttbar$ mass windows for the 1+1 and 1+2 samples.
  The expected SM $\ttbar$ is taken from MC predictions, and
  the expected NTMJ background is derived in Table~\ref{table:mistagRateDerivation}.
}
\label{table:countingExptCrossCheck}
\begin{tabular}{|l|cc|cc| }
\hline
 & \multicolumn{2}{c|}{$m_{\ttbar} = $0.9--1.1\TeVcc} & \multicolumn{2}{c|}{$m_{\ttbar} = $1.3--2.4\TeVcc} \\
                                 &  1+1          &  1+2          &  1+1          &  1+2  \\
\hline
Expected SM $\ttbar$ events      &   69 $\pm$ 36 &  110 $\pm$ 62 &   65 $\pm$ 42 &   24 $\pm$ 15 \\
Expected non-top multijet events &  443 $\pm$ 23 & 1239 $\pm$ 32 &  741 $\pm$ 32 &  817 $\pm$ 38  \\
Total expected events            &  512 $\pm$ 43 & 1349 $\pm$ 70 &  806 $\pm$ 53 &  841 $\pm$ 41 \\
Observed events                  &  506          & 1383          &  809          &  841  \\
\hline
\end{tabular}
\end{center}
\end{table}

\section{Systematic uncertainties}
\ifnpas
\section{Systematic uncertainties}
\fi
\label{sec:systematics}
The sources of systematic uncertainty on the $\ttbar$ invariant mass spectrum
fall into three categories:
(i) the determination of the efficiency,
(ii) the mistag probability,
and (iii) the shape of the $\ttbar$ invariant-mass distribution.
Several sources of systematic uncertainties can simultaneously affect
these three categories, and in such cases, any changes in parameters
have to be varied in a correlated way.
The uncertainties on efficiency include uncertainties
in the overall jet-energy scale for tagged jets ($\approx 2$--$4$\% from
standard jet energy corrections, $\approx 3$\% from the application of the
jet corrections to pruned jets, and $\approx 1$\% to account for the
uncertainty on the determination of the $\wboson$ mass in data and MC,
as described in Sec.~\ref{sec:signalestimate}),
integrated luminosity (2.2\%),
subjet-selection efficiency ($\approx 6$\%, as described in
Sec.~\ref{sec:signalestimate}),
jet energy resolution ($< 1\%$),
and jet angular
resolution ($< 1\%$).
The trigger uncertainty for 1+1 events is 13\% for $m_{\ttbar} = 1$\TeVcc,
and $< 1\%$ for $\ttbar$ masses above 1.5\TeVcc.
The trigger uncertainty is larger for 1+2 events:
20\% for $m_{\ttbar} = 1$\TeVcc,
and 3\% for $m_{\ttbar} > 1.5$\TeVcc, as described in Sec.~\ref{sec:samples}.
The impact of changes in parton distribution functions~\cite{pdf4lhc} is
found to be negligible.

Similar uncertainties affect the $\ttbar$
continuum background and are estimated in the same manner.
In addition, the large uncertainty on the renormalization and
factorization scales (a factor of two)
is found to have significant impact on SM $\ttbar$ production,
resulting in a 50\% variation in the yield, estimated
from MC studies.
This is reflected in the uncertainties on the number of $\ttbar$ events in
Table~\ref{table:countingExptCrossCheck}.
Table~\ref{table:effisystematics} provides a summary of this
information.

\begin{table}[h*]
\begin{center}
\caption{Summary of relative systematic uncertainties on
signal efficiency for two $\ttbar$ mass windows. All values
are in percent. The central value of the subjet selection scale
factor is 0.94, it is the only scale factor that has a non-unit mean.}
\label{table:effisystematics}
\begin{tabular}{|l|c|cc|cc|}
\hline
Source & Variation & \multicolumn{2}{c|}{$m_{\ttbar} = $0.9--1.1\TeVcc} & \multicolumn{2}{c|}{$m_{\ttbar} = $1.3--2.4\TeVcc} \\
                   &          &  1+1          &  1+2          &  1+1          &  1+2  \\
\hline
MC Statistical                          &                 & 2.0  & 1.6  & 0.7  & 1.6 \\
Trigger                                 & See text        & 13   & 20   & $<$1 & 3   \\
Jet energy scale                        & $\approx \pm 5$ & 19   & 19   & 2    & 2   \\
Subjet efficiency scale factor          &    $\pm$ 6      &  6   & 6    & 6    & 6   \\
Luminosity                              &  $\pm$ 2.2      & 2.2  & 2.2  & 2.2  & 2.2 \\
\hline
Total                                   &                 &  24  & 28   &  7   & 8   \\
\hline
\end{tabular}
\end{center}

\end{table}

The uncertainties on the mistagged NTMJ background
include the statistical uncertainty on the sample after loose selection, to
which the mistag probability is applied; the statistical uncertainty
on the mistag probability
itself,
ranging from $< 1\%$ at 1\TeVcc to $\approx 10\%$ at 3\TeVcc;
and the systematic uncertainty on the mistag probability application,
as described in Sec.~\ref{sec:backgroundestimate}, which is in the
range of 1 to 5\% depending on the $\ttbar$ mass.
The total background uncertainty is $\approx 5$\% for the
low-mass region, dominated by the systematic uncertainty, and $\approx
100$\%
for the high-mass region, dominated by statistical
uncertainty associated with the sample after loose selection.

\ifnpas
There may be additional systematic effects in the subjet quantities
that are not present in the hard jet. These systematic effects are
investigated below and summarized in Table~\ref{table:effisystematics}.
In order to estimate the uncertainty due to hard jet and subjet energy
scales, the energy is increased and decreased, and the
effect is checked in the Monte Carlo. This effect
is also propagated to the full invariant mass spectrum in a correlated
way.
The hard jet energy resolution is already known to be slightly worse
than in the simulation, so an additional smearing affect is added
to our jets as per the recommendation of the JME POG~\cite{jetid}.
Additionally, further subjet energy and angular
resolution effects are investigated below.

\subsection {Signal efficiency uncertainties}

\noindent
To determine the actual number of signal events in the window,
the number of events from the Monte Carlo is scaled by
several data-to-Monte-Carlo scale factors.
\begin{itemize}
\item Trigger, determined by weighting the
  $\mathrm{Z}'$ Monte Carlo by a trigger efficiency derived
  from Monte Carlo.
  The uncertainty is taken as half the difference between the weighted
  and unweighted distributions in the Monte Carlo, cross checked as
  the difference in efficiency after tagging between data and Monte Carlo.
  This depends on the mass window, but as an example, the 1\TeVcc mass
  window for the 1+2 analysis (where the effect is largest) is
  $C_{trig} = 0.65 \pm 0.17$, from Section~\ref{sec:datasampleAndSelection}.
\item Jet energy scale, determined by checking the $\mathrm{W}$ boson mass
  reconstructed by the jet pruning algorithm in a
  semileptonic $\mathrm{t}\overline{\mathrm{t}}$
  sample. The value is consistent with unity with uncertainty 5\%. This is
  added in quadrature to the uncertainties from the standard
  anti-$k_{\mathrm T}$ 0.5
  jet energy corrections ($\sigma_{AK5}$, which is a relative uncertainty).
  The value used for the scale factor is
  $C_{JES} = (1.00 \pm 0.05) \otimes (C_\text{AK5} \pm \sigma_{AK5})$ (where
  $C_\text{AK5} \pm \sigma_{AK5}$ is the \texttt{L2L3Residual} correction and
  uncertainty), from Section~\ref{sec:substructure_eff}.
\item Substructure tagging, determined by checking the ratio of efficiencies
  of $\mathrm{W}$ tagging in a semileptonic $\mathrm{t}\overline{\mathrm{t}}$
  sample. The scale factor is $C_{tag} = \scalefactor$,
  from Section~\ref{sec:substructure_jec}.
\end{itemize}

\noindent
After these corrections, the estimated number of $\mathrm{Z}'$ events in
the analysis is therefore

\begin{equation}
N_{\zp}=N_{\zp}^{MC} \times C_{tag}^{2} \times C_{JES} \times C_{trig}
\end{equation}
where $N_{\zp}^{MC}$ is the number of $\mathrm{Z}'$ events in the Monte
Carlo and the $C_X$ are defined above.

\vspace{4mm}
\noindent
The relative uncertainty on the signal is therefore

\begin{equation}
\frac{\sigma_{\zp} }{N_{\zp}} =  \sqrt{\left(\frac{\sigma_{{\zp}^{MC}}}{N_{{\zp}^{MC}}}\right)^{2}
 + \left(2\times\frac{\sigma_{C_{tag}}}{C_{tag}}\right)^{2}
 + \left(\frac{\sigma_{C_{JES}}}{C_{JES}}\right)^{2}
 + \left(\frac{\sigma_{C_{trig}}}{C_{trig}}\right)^{2}
 + \left(\frac{\sigma_{C_{lumi}}}{C_{lumi}}\right)^{2}
 + \left(\frac{\sigma_{C_{Xsection}}}{C_{Xsection}}\right)^{2}
 }
 \end{equation}

Here $\sigma_{{\zp}^{MC}}$ is the Monte Carlo statistical error,
$\sigma_{C_{tag}}$ is the tagging scale factor uncertainty,
$\sigma_{C_\text{JEC}}$ is the the jet energy correction uncertainty, and
$\sigma_{C_\text{trig}}$ is the trigger weighting uncertainty.

The trigger uncertainty is determined by measuring half the difference
between the number of events in the signal window in the trigger weighted
histogram and the number of events in the non-weighted histogram:

\begin{eqnarray}
C_\text{trig} & = & \frac{N_{Z'}^{MC} \mathrm{ (Weighted)}} {N_{Z'}^{MC} \mathrm{ (Unweighted)}} \\
\frac{\sigma_{C_\text{trig}}}{C_\text{trig}} & = & \frac{1}{2} \times \left|1 - C_\text{trig}\right|
\label{trig}
\end{eqnarray}

\vspace{4 mm}
\noindent
The jet pT was scaled up and down in order to determine the JEC error. The ratio of signal window efficiency $\epsilon_{\text{window}}=N_{\text{window}}/N_{\text{selected}}$ before and after scaling was measured by Sal (see http://hep.pha.jhu.edu:8080/boostedtop/1381 for signal MC and http://hep.pha.jhu.edu:8080/boostedtop/1402 for top MC):

\begin{equation}
R_{\text{up}}=\frac{\epsilon_{\text{scale pT up}}}{\epsilon_{\text{no scale}}}  \quad \quad  \quad \quad R_{\text{down}}=\frac{\epsilon_{\text{scale pT down}}}{\epsilon_{\text{no scale}}}
\end{equation}

\noindent
The uncertainty in the number of events in the signal window from JEC uncertainties is given by

\begin{equation}
\frac{\sigma_{C_\text{JEC}}}{C_\text{JEC}}  =  \mathrm{max}( | 1 - R_{up}|, |1 - R_{down}|)
\label{jec}
\end{equation}

\vspace{4 mm}

\noindent
The efficiency is the fraction of events in the signal window.

\begin{eqnarray}
N_{window}(m_0) &=& \sum_{m_0 - \Delta_-}^{m_0 + \Delta_+} m_{\mathrm{t}\overline{\mathrm{t}}} \\
N_{total} &=& { \sum m_{\mathrm{t}\overline{\mathrm{t}}} }  \\
\epsilon & = & \frac{N_{window}}{N_{total}}
\end{eqnarray}
\vspace{4 mm}

\noindent
\subsection {Background estimation uncertainties}
\noindent
The number of background events in the signal window is the sum of the NTMJ prediction and the $\ttbar$ MC:

\begin{equation}
N_{\text{BKG}} =  N_{\text{NTMJ}}+N_{{\ttbar}}
\end{equation}
\vspace{2mm}

The uncertainties on the $\mathrm{t}\overline{\mathrm{t}}$ are derived in the
same way as the signal Monte Carlo.

\begin{equation}
N_{\ttbar}=N_{\ttbar}^{MC} \times C_{tag}^{2} \times C_{JES} \times C_{trig}
\end{equation}
where $N_{\ttbar}^{MC}$ is the number of $\ttbar$ events in the Monte
Carlo and the $C_X$ are defined above.

\vspace{4mm}
\noindent
The relative uncertainty on the $\ttbar$ background is therefore

\begin{equation}
\label{eq:dttbar}
\frac{\sigma_{\ttbar} }{N_{\ttbar}} =  \sqrt{\left(\frac{\sigma_{\ttbar}^{MC}}{N_{\ttbar}^{MC}}\right)^{2}
 + \left(2\times\frac{\sigma_{C_{tag}}}{C_{tag}}\right)^{2}
 + \left(\frac{\sigma_{C_{JES}}}{C_{JES}}\right)^{2}
 + \left(\frac{\sigma_{C_{trig}}}{C_{trig}}\right)^{2}
 + \left(\frac{\sigma_{C_{lumi}}}{C_{lumi}}\right)^{2}
 + \left(\frac{\sigma_{C_{Xsection}}}{C_{Xsection}}\right)^{2}
 }
\end{equation}

The number of NTMJ events is determined by weighting the {\it sideband} sample
(single-tagged events in the 1+1 case, pretagged events in the 1+2
case) by the probability to tag the type 1 jet with the mistag probability
derived in Section~\ref{sec:mistag_rate} (the weighting procedure is denoted by
``$\otimes$''). The mistag probability is a function of the jet's transverse
momentum.

\begin{equation}
N_\text{NTMJ} = N_\text{pretag} \otimes P_m(p_T)
\end{equation}

\noindent
The error on the number of events from the NTMJ prediction is given by:

\begin{equation}
\label{eq:dqcd}
\frac{\sigma_{\text{NTMJ}}}{ N_\text{NTMJ}} = \sqrt{ \left(\frac{\sigma_{N_\text{pre}}}{N_\text{pre}}\right)^2 \oplus \left(\frac{\sigma_{P_m}}{P_m}\right)^2 }
\end{equation}
where ``$\oplus$'' denotes the uncertainty propagation scheme of the weighting
procedure.

\vspace{8 mm}
\noindent
Here  $\sigma_{P_m}$ is a systematic uncertainty which takes into account the different mistag measurements.
The nominal background estimations use the mass-modified-mistag procedure.
The systematic uncertainty assigned to this procedure is estimated as half
the difference between the mass-modified-mistag approach ($P_m$) and the
{\it standard mistag} approach ($P_\text{standard}$).

\begin{equation}
\frac{\sigma_{P_{m}}}{P_m}  = \frac{1}{2} \left| 1 - \frac{P_\text{standard}}{P_m} \right|
\end{equation}

\noindent
The error on the number of background events is thus given by

\begin{equation}
\sigma_{\text{BKG}} = \sqrt{\sigma_{\text{NTMJ}}^{2} +  \sigma_{\ttbar}^{2}}
\end{equation}
where $\sigma_{\text{NTMJ}}$ is given by Equation~\ref{eq:dqcd} and $\sigma_{\ttbar}$
is given by Equation~\ref{eq:dttbar}.

The systematic effects are now described
in detail.

Table~\ref{table:bkgsystematics} summarizes the relative systematic uncertainties on the background
estimation $N_{B}$.

\begin{table}[h]
\begin{center}
\begin{tabular}{|l|r|r|r|r|r|}
\hline
Resonance Mass                           &                & M=1   & M=1.5    & M=2.0    & M=3.0   \\
\hline
Mass Window                              &                & 0.9-1.1 & 1.2-1.6 & 1.3-2.4 & 2.0-3.3 \\
\hline
Source                                   & Variation      & \multicolumn{4}{c|}{Relative Effect (All values in percent)} \\
\hline
\hline
\multicolumn{6}{|c|}{Type 1+1} \\
\hline
\multicolumn{6}{|c|}{\ttbar Monte Carlo} \\
Monte Carlo Statistics                   & N/A            & 5     & 5     & 5     & 26    \\
JEC                                      & $\approx \pm 6$& 3     & 28    & 40    & 53    \\
Trigger                                  & See text       & 14    & 2     & 1     & $<$1  \\
Substructure tagging scale factor        & 90 $\pm$ 4     & 9     & 9     & 9     & 9     \\
Luminosity       		  	             & $\pm$ 4.5      & 4.5   & 4.5   & 4.5   & 4.5     \\
$\ttbar$ cross section                   & $\pm$ 10       & 10    & 10    & 10    & 10    \\
Total                                    &                & 21    & 32    & 43    & 61    \\
\hline
\multicolumn{6}{|c|}{NTMJ} \\
NTMJ prediction statistical               & N/A            & 1     & 1     & 1     & 3    \\
Mistag kinematics                        & See text       & 5     & 4     & 4     & 2    \\
Total                                    &                & 5     & 4     & 4     & 4    \\
\hline
\multicolumn{6}{|c|}{Total} \\
\ttbar weight                            &                &  11   &  9   &    6  &    3  \\
NTMJ weight                               &                &   89  &  91   &   94  &   97  \\
Total background                         &                & 5     &   5   &  5    & 4  \\
\hline
\hline
\multicolumn{6}{|c|}{Type 1+2} \\
\hline
\multicolumn{6}{|c|}{\ttbar Monte Carlo} \\
Monto Carlo Statistics                   &  N/A           & 4     & 7     & 9     &  71    \\
JEC                                      & $\approx \pm 6$& 3     & 28    & 40    & 53    \\
Trigger                                  & See text       & 24    & 8     & 7     & $<$1   \\
Substructure tagging scale factor        & 90 $\pm$ 4     & 9     & 9     & 9     & 9     \\
Luminosity       			             & $\pm$ 4.5      & 4.5   & 4.5   & 4.5   & 4.5   \\
$\ttbar$ cross section                   & $\pm$ 10       & 10    & 10    & 10    & 10    \\
Total                                    &                & 28    & 33    & 44    & 90    \\
\hline
\multicolumn{6}{|c|}{NTMJ} \\
NTMJ prediction statistical               & N/A            & 1     & 1     & 1     & 3    \\
Mistag kinematics                        & See text       & 3     & 6     & 5     &  2    \\
Total NTMJ prediction                     &                & 3     & 6     & 5     & 3    \\
\hline
\multicolumn{6}{|c|}{Total} \\
\ttbar weight                            &                &   7  &    4  &    3  & $<1$  \\
NTMJ weight                               &                &  93  &   96  &   97  &  100  \\
Total background                         &                &   3  &  6    & 5     & 3    \\
\hline
\end{tabular}
\end{center}
\caption{Summary of statistical and systematic relative uncertainties for the background estimation in a mass window.
The uncertainty for the mistag kinematics is taken as half the difference between the procedure when the mass-modified-mistag
probability is used, and if the observed jet mass is used. The uncertainty for the trigger selection is taken as half the
difference between a Monte-Carlo-derived trigger efficiency weighting scheme, and the raw Monte Carlo efficiency.
This is also cross-checked in the data, and the uncertainties assigned
cover the differences observed. All values are in percent.}
\label{table:bkgsystematics}
\end{table}

\subsection{Jet energy scale}

The subjet energy scale is derived for the jet pruning algorithm in
Section~\ref{sec:substructure_jec}. An additional 5\% uncertainty
is measured for the subjet energy scale, which is added in quadrature
to the uncertainties of the overall jet energy corrections described
in Section~\ref{sec:reconstruction}. We thus vary the four momenta
of the jets and subjets up and down by this quadrature-summed amount
and investigate the effect on the final result.
Figure~\ref{figs:jes_sys} shows the
systematic uncertainty due to this effect.
Since we are performing a counting experiment, the relevant effect is
the change in the expected number of events in the signal window.
The relative change in event yields is then taken as a systematic
uncertainty.
The jet energy scale uncertainties
in Tables~\ref{table:effisystematics} and \ref{table:bkgsystematics}
have
a nontrivial dependence on the invariant mass of the postulated
resonance because of these effects. For the first mass point (1
\TeVcc), the signal window is rather small, and hence the effect of
changing the jet energy scale is large. As the mass is increased this
effect lessens. However, at the very last mass point (3 \TeVcc), the
uncertainty rises again, this time due to the radiative tail
of the distribution seen in Figure~\ref{figs:type11ttmass}. Changes
in the jet energy scale impact the tail of the distribution
differently than the core, and hence the effect is magnified.

\begin{figure}
\centering
\includegraphics[width=0.7\textwidth]{figs/Zprime_M1000GeV_W10GeV_Type11Analyzer_Scale}
\caption{Jet energy scale uncertainty. The effect of
  this systematic uncertainty on the signal efficiency and shape is
  the dominant signal systematic uncertainty.}
\label{figs:jes_sys}
\end{figure}

\subsection{Jet energy resolution}

The jet energy in the simulation is slightly too optimistic,
at around the 10$\pm$10\% level. Thus, we smear the jet energies
in the simulation by a factor of 10\% for the central value, and
use 0\% and 20\% as uncertainties.
At the present time, the subjet angular resolution
is taken to be the same as the hard jet and are varied
with that assumption.
There is no significant effect.

Figures~\ref{figs:Zprime_M1000GeV_W10GeV_Type11Analyzer_PtSmear}-
\ref{figs:Zprime_M3000GeV_W30GeV_Type12Analyzer_PtSmear}
show the effect of the jet energy smearing for the other $\zp$ masses,
as well as for the Type 1+2 analysis.

At the present time, the subjet energy resolution
is taken to be the same as the hard jet and are varied
with that assumption. They are included in Figure~\ref{figs:jer_sys}.

\subsection{Jet angular resolution}

Similar to the jet energy resolution, the jet angular resolution is
also investigated. The variation of the subjet angular resolution is
taken to be a variation of 0\%, 10\%, and 20\%.
At the present time, the subjet angular resolution
is taken to be the same as the hard jet and are varied
with that assumption.
There is no significant effect.

Figures~\ref{figs:Zprime_M1000GeV_W10GeV_Type11Analyzer_EtaSmear}-
\ref{figs:Zprime_M3000GeV_W30GeV_Type12Analyzer_EtaSmear}
show the effect of the jet energy smearing for the various $\zp$ masses,
for the Type 1+1 and 1+2 analyses.
They are included in Figure~\ref{figs:jar_sys}.

\begin{figure}
\centering
\includegraphics[width=0.65\textwidth]{figs/Zprime_M1000GeV_W10GeV_Type11Analyzer_PtSmear}
\caption{Jet energy resolution smearing uncertainty for a $\zp$ with mass 1 \TeVcc
  and width 10\GeVcc, for the Type 1+1 analysis.}
\label{figs:jer_sys}
\label{figs:Zprime_M1000GeV_W10GeV_Type11Analyzer_PtSmear}
\end{figure}

\begin{figure}
\centering
\includegraphics[width=0.65\textwidth]{figs/Zprime_M2000GeV_W20GeV_Type11Analyzer_Scale}
\caption{Jet energy scale uncertainty for a $\zp$ with mass 2 \TeVcc
  and width 20\GeVcc, for the Type 1+1 analysis.}
\label{figs:Zprime_M2000GeV_W20GeV_Type11Analyzer_Scale}
\end{figure}

\begin{figure}
\centering
\includegraphics[width=0.65\textwidth]{figs/Zprime_M3000GeV_W30GeV_Type11Analyzer_Scale}
\caption{Jet energy scale uncertainty for a $\zp$ with mass 3 \TeVcc
  and width 30\GeVcc, for the Type 1+1 analysis.}
\label{figs:Zprime_M3000GeV_W30GeV_Type11Analyzer_Scale}
\end{figure}

\begin{figure}
\centering
\includegraphics[width=0.65\textwidth]{figs/Zprime_M1000GeV_W10GeV_Type12Analyzer_Scale}
\caption{Jet energy scale uncertainty for a $\zp$ with mass 1 \TeVcc
  and width 10\GeVcc, for the Type 1+2 analysis.}
\label{figs:Zprime_M1000GeV_W10GeV_Type12Analyzer_Scale}
\end{figure}

\begin{figure}
\centering
\includegraphics[width=0.65\textwidth]{figs/Zprime_M2000GeV_W20GeV_Type12Analyzer_Scale}
\caption{Jet energy scale uncertainty for a $\zp$ with mass 2 \TeVcc
  and width 20\GeVcc, for the Type 1+2 analysis.}
\label{figs:Zprime_M2000GeV_W20GeV_Type12Analyzer_Scale}
\end{figure}

\begin{figure}
\centering
\includegraphics[width=0.65\textwidth]{figs/Zprime_M3000GeV_W30GeV_Type12Analyzer_Scale}
\caption{Jet energy scale uncertainty for a $\zp$ with mass 3 \TeVcc
  and width 30\GeVcc, for the Type 1+2 analysis.}
\label{figs:Zprime_M3000GeV_W30GeV_Type12Analyzer_Scale}
\end{figure}

\begin{figure}
\centering
\includegraphics[width=0.65\textwidth]{figs/Zprime_M1000GeV_W10GeV_Type11Analyzer_EtaSmear}
\caption{Jet angular resolution smearing uncertainty for a $\zp$ with mass 1 \TeVcc
  and width 10\GeVcc, for the Type 1+1 analysis.}
\label{figs:jar_sys}
\label{figs:Zprime_M1000GeV_W10GeV_Type11Analyzer_EtaSmear}
\end{figure}

\begin{figure}
\centering
\includegraphics[width=0.65\textwidth]{figs/Zprime_M2000GeV_W20GeV_Type11Analyzer_PtSmear}
\caption{Jet energy resolution smearing uncertainty for a $\zp$ with mass 2 \TeVcc
  and width 20\GeVcc, for the Type 1+1 analysis.}
\label{figs:Zprime_M2000GeV_W20GeV_Type11Analyzer_PtSmear}
\end{figure}

\begin{figure}
\centering
\includegraphics[width=0.65\textwidth]{figs/Zprime_M3000GeV_W30GeV_Type11Analyzer_PtSmear}
\caption{Jet energy resolution smearing uncertainty for a $\zp$ with mass 3 \TeVcc
  and width 30\GeVcc, for the Type 1+1 analysis.}
\label{figs:Zprime_M3000GeV_W30GeV_Type11Analyzer_PtSmear}
\end{figure}

\begin{figure}
\centering
\includegraphics[width=0.65\textwidth]{figs/Zprime_M1000GeV_W10GeV_Type12Analyzer_PtSmear}
\caption{Jet energy resolution smearing uncertainty for a $\zp$ with mass 1 \TeVcc
  and width 10\GeVcc, for the Type 1+2 analysis.}
\label{figs:Zprime_M1000GeV_W10GeV_Type12Analyzer_PtSmear}
\end{figure}

\begin{figure}
\centering
\includegraphics[width=0.65\textwidth]{figs/Zprime_M2000GeV_W20GeV_Type12Analyzer_PtSmear}
\caption{Jet energy resolution smearing uncertainty for a $\zp$ with mass 2 \TeVcc
  and width 20\GeVcc, for the Type 1+2 analysis.}
\label{figs:Zprime_M2000GeV_W20GeV_Type12Analyzer_PtSmear}
\end{figure}

\begin{figure}
\centering
\includegraphics[width=0.65\textwidth]{figs/Zprime_M3000GeV_W30GeV_Type12Analyzer_PtSmear}
\caption{Jet energy resolution smearing uncertainty for a $\zp$ with mass 3 \TeVcc
  and width 30\GeVcc, for the Type 1+2 analysis.}
\label{figs:Zprime_M3000GeV_W30GeV_Type12Analyzer_PtSmear}
\end{figure}

\begin{figure}
\centering
\includegraphics[width=0.65\textwidth]{figs/Zprime_M2000GeV_W20GeV_Type11Analyzer_EtaSmear}
\caption{Jet angular resolution smearing uncertainty for a $\zp$ with mass 2 \TeVcc
  and width 20\GeVcc, for the Type 1+1 analysis.}
\label{figs:Zprime_M2000GeV_W20GeV_Type11Analyzer_EtaSmear}
\end{figure}

\begin{figure}
\centering
\includegraphics[width=0.65\textwidth]{figs/Zprime_M3000GeV_W30GeV_Type11Analyzer_EtaSmear}
\caption{Jet angular resolution smearing uncertainty for a $\zp$ with mass 3 \TeVcc
  and width 30\GeVcc, for the Type 1+1 analysis.}
\label{figs:Zprime_M3000GeV_W30GeV_Type11Analyzer_EtaSmear}
\end{figure}

\clearpage

\begin{figure}
\centering
\includegraphics[width=0.65\textwidth]{figs/Zprime_M1000GeV_W10GeV_Type12Analyzer_EtaSmear}
\caption{Jet angular resolution smearing uncertainty for a $\zp$ with mass 1 \TeVcc
  and width 10\GeVcc, for the Type 1+2 analysis.}
\label{figs:Zprime_M1000GeV_W10GeV_Type12Analyzer_EtaSmear}
\end{figure}

\begin{figure}
\centering
\includegraphics[width=0.65\textwidth]{figs/Zprime_M2000GeV_W20GeV_Type12Analyzer_EtaSmear}
\caption{Jet angular resolution smearing uncertainty for a $\zp$ with mass 2 \TeVcc
  and width 20\GeVcc, for the Type 1+2 analysis.}
\label{figs:Zprime_M2000GeV_W20GeV_Type12Analyzer_EtaSmear}
\end{figure}

\begin{figure}
\centering
\includegraphics[width=0.65\textwidth]{figs/Zprime_M3000GeV_W30GeV_Type12Analyzer_EtaSmear}
\caption{Jet angular resolution smearing uncertainty for a $\zp$ with mass 3 \TeVcc
  and width 30\GeVcc, for the Type 1+2 analysis.}
\label{figs:Zprime_M3000GeV_W30GeV_Type12Analyzer_EtaSmear}
\end{figure}

\subsection{Substructure Selection Efficiency}

The efficiency of the substructure selections is outlined above,
in Section~\ref{sec:substructure_eff}. Since we are using two substructure
tags in both the 1+1 and 1+2 analyses, the uncertainty on
the signal efficiency (and the \ttbar background Monte Carlo estimate)
is twice the efficiency derived in Section~\ref{sec:substructure_eff}.
Thus, the scale factor due to substructure selection efficiency
for the analysis is \scalefactorsqrd.

\subsection{Trigger Efficiency}

The trigger efficiency impacts the \ttbar continuum background,
as well as the \zp signal samples. The trigger efficiency is
determined from Monte Carlo, and cross-checked using a data-driven
technique.
The difference
between MC and data is approximately a half of the trigger inefficiency
estimated in MC. Half of the trigger inefficiency is taken as an
estimate of the systematic uncertainty.

\fi

\section{Statistical treatment}

\label{sec:statistics_paper}

The main result of this analysis is the fit to data assuming a
resonance hypothesis for the new
physics, in which a
likelihood is fit to the expected $\ttbar$ invariant mass
distributions for signal and background.
The second result corresponds to a counting of events relative to some
generic model of an enhancement of the $\ttbar$ continuum assuming the
SM efficiency for the additional contribution.
These two results are discussed below.

\subsection{Resonance analysis}
\label{sec:resonant}

The first analysis uses a resonant signal hypothesis to search for localized
contributions to the
$m_{\ttbar}$ spectrum.
A cross-check of the analysis is performed by counting the number of
events in a mass range defined for each resonant mass value.
Two such mass ranges are shown in Tables~\ref{table:mistagRateDerivation}-\ref{table:effisystematics}.
In the main resonance analysis and the cross-check, the number of observed
events, $N_{obs}$, is
compared to the expectation $N_{exp}$, based on the production cross section
$\sigma_{\zp}$, branching fraction $B(\zp \to \ttbar)$, signal reconstruction
efficiency $\epsilon$, integrated luminosity
$L$,  
and the predicted number of background events $N_B$:

\begin{equation}\label{eq:N_exp_signal_bkg}
N_{exp} = \sigma_{\zp} \times B(\zp \to \ttbar)
         \times \epsilon  \times  L +  N_B.
\end{equation}

The likelihood is computed using the Poisson probability to
observe $N_{obs}$, given a mean of $N_{exp}$, with uncertain parameters
$L$, $\epsilon$, and $N_B$, all defined through
log-normal priors based on their mean values and their uncertainties.
The shapes and normalizations of signal and background distributions
are varied within their systematic uncertainties until the likelihood
is maximized. This procedure effectively integrates over the parameters
describing the systematic uncertainties, thereby reducing their
impact.

The upper
limits at $95\%$ confidence level (CL)
on the product of the \zp production cross section and the branching fraction to the $\ttbar$ final
state are extracted for the combination of the 1+1 and
1+2 $\ttbar$ mass spectra, as a function of $\mzp$ in a range from 1
to 3\TeVcc with a 0.1\TeVcc increment.
A CL$_S$ method~\cite{Read1,junkcls,cls_lhc} is used to extract the
95\% CL\ upper limits, with the posterior based on a Poisson model for each bin of the
$m_{\ttbar}$ distribution.

\begin{figure}[htbp]
\begin{center}
\subfloat[]{\label{figs:limit_shape_1percent}\includegraphics[width=0.60\textwidth]{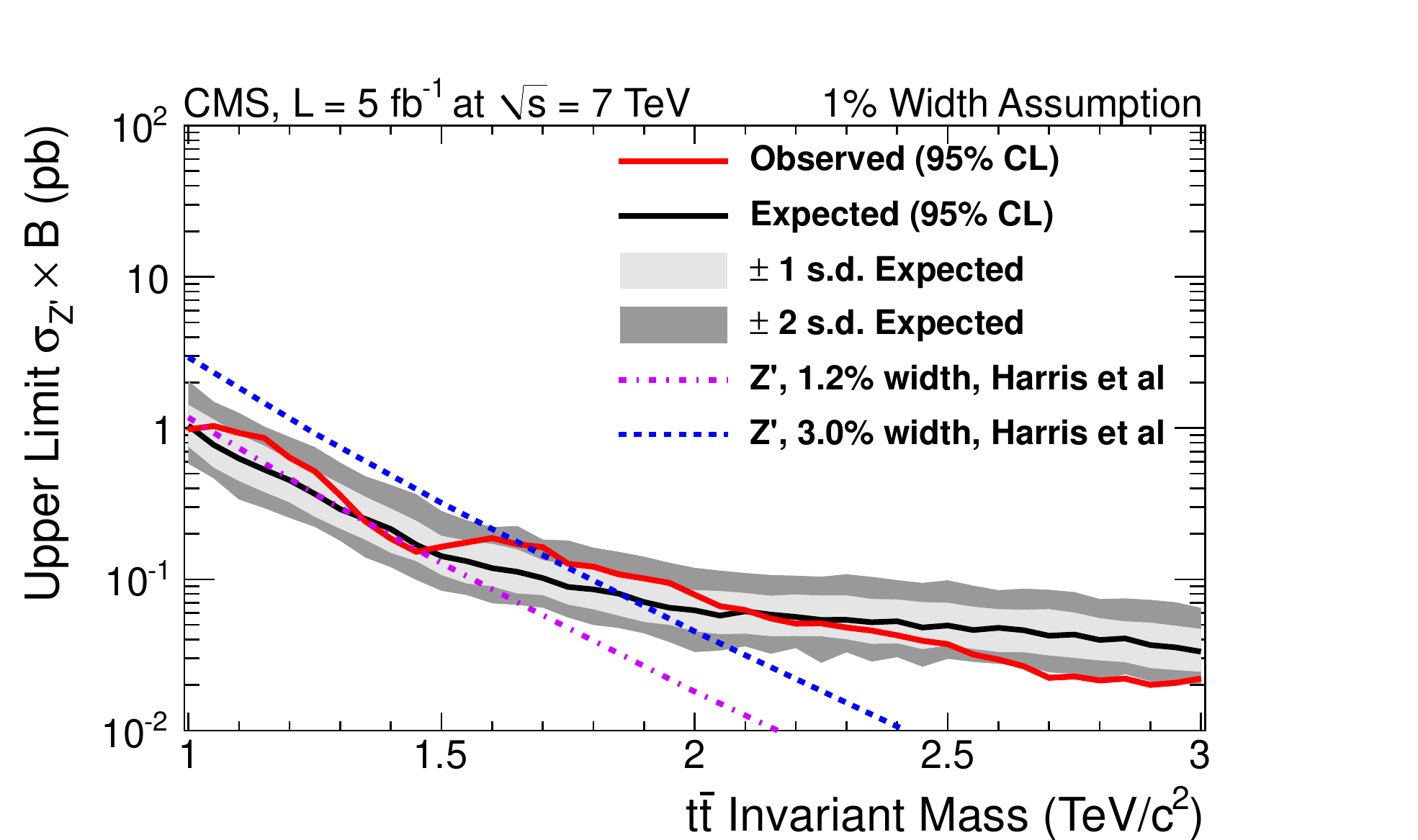}}\\
\subfloat[]{\label{figs:limit_shape_10percent}\includegraphics[width=0.6\textwidth]{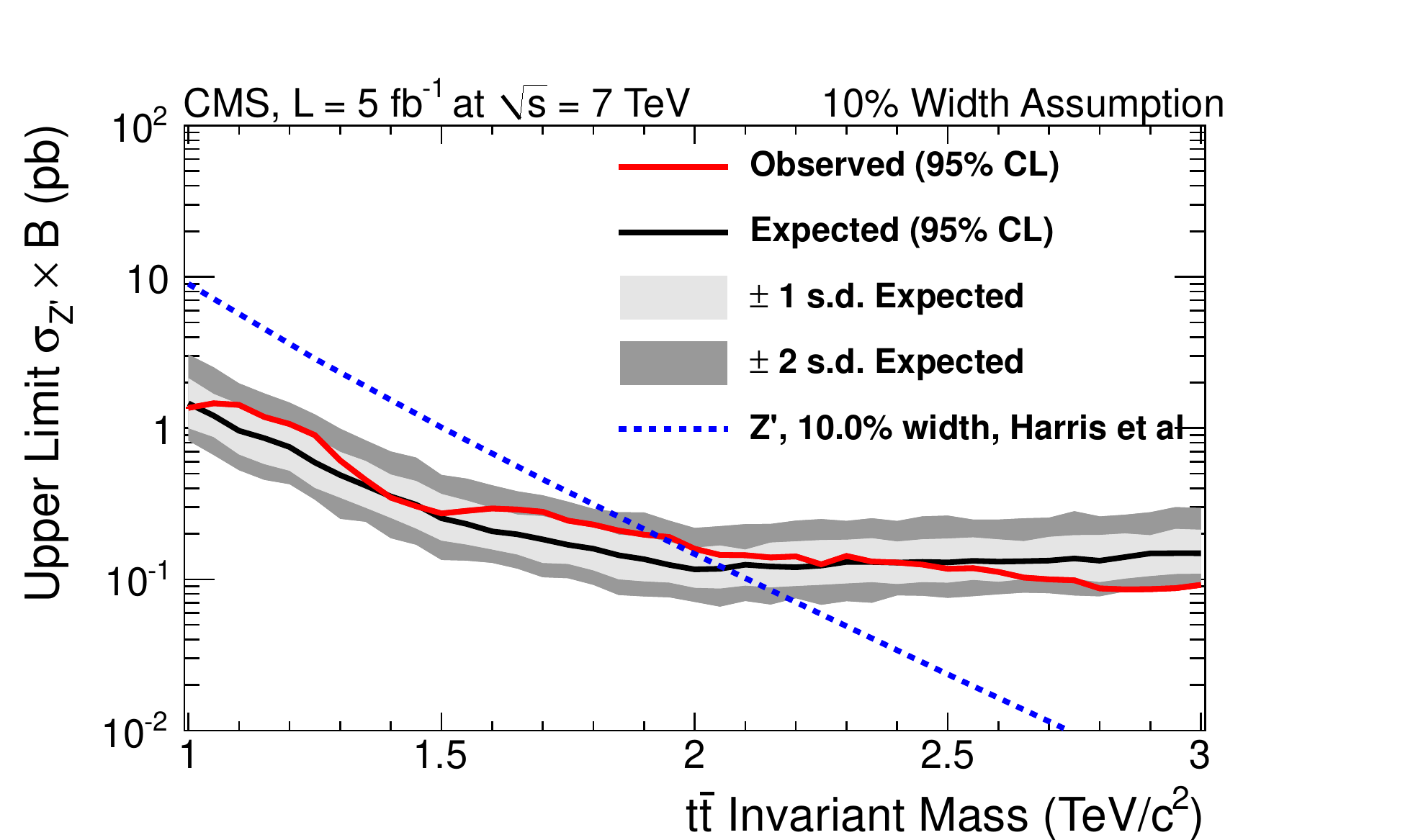}}\\
\subfloat[]{\label{figs:limit_shape_kkg}\includegraphics[width=0.60\textwidth]{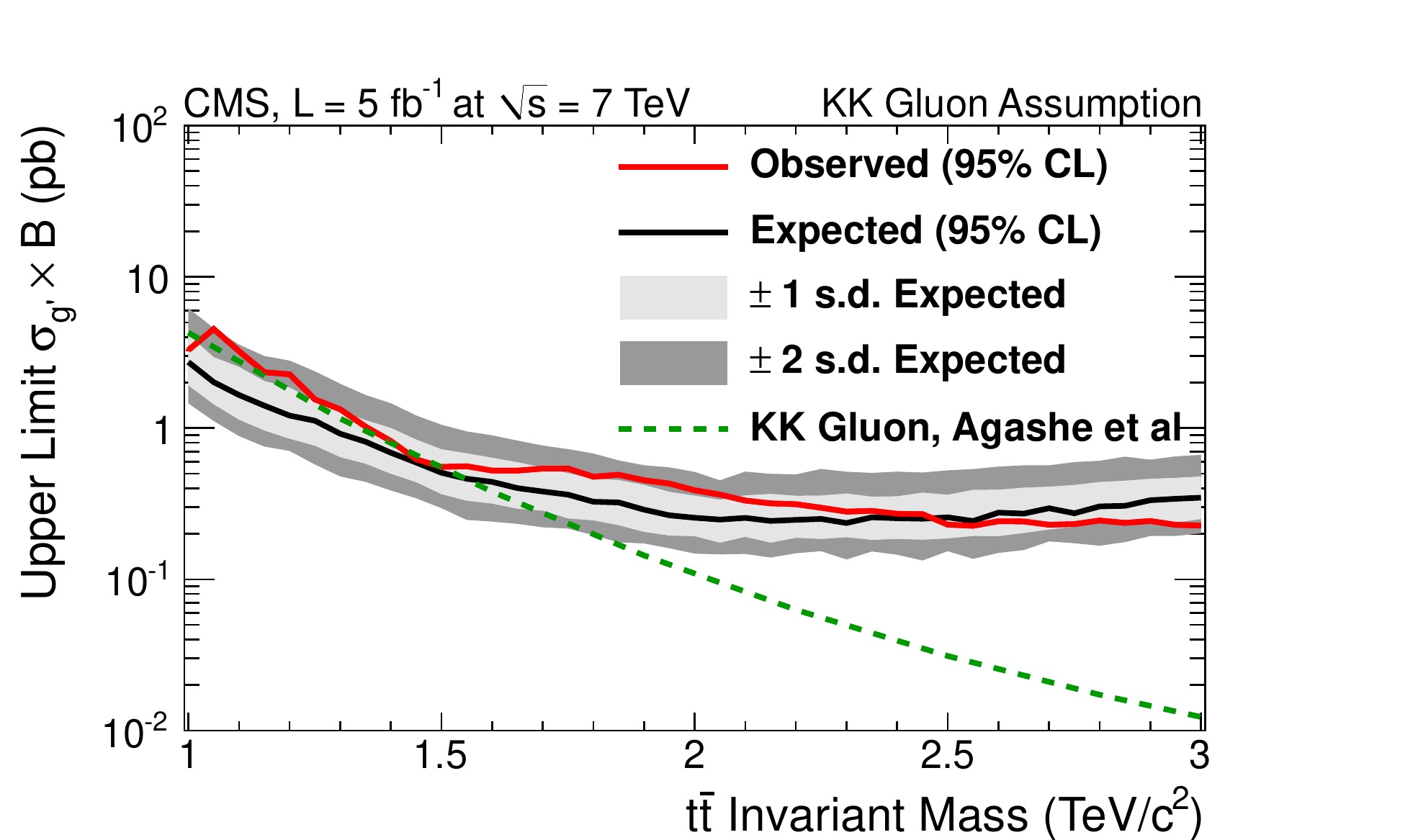}}
\caption{The 95\% CL upper limits on the product of production cross
  section ($\sigma$) and branching fraction ($B$) of hypothesized objects into $\ttbar$, as a
  function of assumed resonance
  mass. (a) $\zp$ production with $\Gamma_{\zp} / \mzp = $ 1\% (1\% width assumption) compared to
  predictions based on Refs.~\cite{nsd3,Hill:1991at,Hill:1994hp} for $\Gamma_{\zp} / \mzp = $1.2\% and 3.0\%.
  (b) $\zp$ production with $\Gamma_{\zp} / \mzp = $ 10\% (10\% width assumption) compared to
  predictions based on Refs.~\cite{nsd3,Hill:1991at,Hill:1994hp} for a width of 10\%.
  (c) Randall--Sundrum Kaluza--Klein gluon production from
  Ref.~\cite{rs_gluon_1}, compared to the theoretical prediction of
  that model. The $\pm 1$ and $\pm 2$ standard deviation (s.d.)
  excursions are shown relative to the results expected for the
  available luminosity.
  }
\label{figs:limit_shape}
\end{center}
\end{figure}

Figure~\ref{figs:limit_shape} shows the observed and expected upper
limits for: (a) a $\zp$
hypothesis with $\Gamma_{\zp} / \mzp = $ 1\%, (b) a $\zp$
hypothesis with $\Gamma_{\zp} / \mzp = $ 10\%, and (c)
a Randall--Sundrum Kaluza--Klein gluon hypothesis. Also
shown are the theoretical predictions for several models to compare to
the observed and expected limits. In
Figure~\ref{figs:limit_shape_1percent}, predictions are also 
shown for a topcolor $\zp$ model based on
Refs.~\cite{nsd3,Hill:1991at,Hill:1994hp} updated to $\sqrt{s} =
7$\TeV in Ref.~\cite{Harris:2011ez}, with
$\Gamma_{\zp} / \mzp = $ 1.2\% and $\Gamma_{\zp} / \mzp = $ 3\%,
compared to limits obtained assuming a 1\% width.
Higher-order QCD corrections to the \zp production cross section were
accounted for through a constant K-factor, computed to be 1.3.
The same $\zp$ model,
but for $\Gamma_{\zp} / \mzp = $ 10\%, is compared
to the limits obtained assuming a 10\% width 
in Fig.~\ref{figs:limit_shape_10percent}. 
Finally, in Fig.~\ref{figs:limit_shape_kkg}, the prediction of the Randall--Sundrum
Kaluza--Klein gluon model from Ref.~\cite{rs_gluon_1} is compared to
the limits from data.

Using the upper limits for the $\zp$ with 1\% width, mass ranges for
two $\zp$ models are excluded as seen in Fig.~\ref{figs:limit_shape_1percent}.
First, the mass range $1.0$--$1.6$\TeVcc is excluded for
a topcolor \zp with width $\Gamma_{\zp} / \mzp =$ 3\%.
Second, two mass ranges, $1.3$--$1.5$\TeVcc and a narrow range
(smaller than the mass increment)
close to $1.0$\TeVcc, are excluded
for the same topcolor \zp with width $\Gamma_{\zp} / \mzp =$ 1.2\%.
Similarly, using the upper limits for the $\zp$ with 10\% width
as seen in Fig.~\ref{figs:limit_shape_10percent},
the mass range $1.0$--$2.0$\TeV
is excluded for a topcolor \zp with width $\Gamma_{\zp} / \mzp =$ 10\%.

Finally, as seen in Fig.~\ref{figs:limit_shape_kkg},
upper limits in the range of 1\unit{pb} are set on
$\sigma_{g'} \times B(g' \to \ttbar)$ for $m_{g'} > 1.4$\TeVcc
for a specific Randall--Sundrum gluon model~\cite{rs_gluon_1}, which
exclude the existence of this particle with masses
between $1.4$--$1.5$\TeVcc, as well as
in a narrow region (smaller than
the mass increment) close to $1.0$\TeVcc.

The resonant analysis is cross-checked by counting
events in specified mass windows of $m_{\ttbar}$.
The signal region is defined by a window in $m_{\ttbar}$, and the
background estimates from Figs.~\ref{figs:ttMassType11LOG}
and \ref{figs:ttMassType12LOG} are integrated over this range.
The results obtained from this cross-check are consistent with the
analysis of the $m_{\ttbar}$ spectrum, but are not as sensitive.
Table~\ref{table:countingExptCrossCheck} shows the number of events
observed and expected in two mass windows
for the 1+1 and 1+2 channels, 0.9--1.1\TeVcc
corresponding to the 1\TeVcc $\zp$ sample, and 1.3--2.4\TeVcc
corresponding to the 2\TeVcc $\zp$ sample.
The
observed 95\% CL upper limits on signal cross section change from 1.0
to 2.0\unit{pb} at 1\TeVcc, from 0.10 to 0.26 pb at 2\TeVcc, and from 0.02
to 0.05\unit{pb} at 3\TeVcc.
Most of the difference is attributed to a better statistical
handling in the resonance analysis of the bins with large background
in the mass distribution.

\subsection{\texorpdfstring{$\ttbar$}{t anti-t} enhancement analysis}
\label{sec:enhance}

In the second analysis, 
general enhancement is assumed in modeling the $\ttbar$ mass spectrum due to
some new phenomenon (NP), assuming the same signal efficiency as
for the SM $\ttbar$ continuum, as described in Refs.~\cite{top_afb_implications1,top_afb_implications2}.
The limit on any possible enhancement
is presented in terms of a variable $\mathcal{S}$, the ratio of the
integral of the $m_{\ttbar}$ distribution above 1\TeVcc corresponding
to SM $\ttbar$ production and a contribution from some NP, to that from
just SM $\ttbar$ production:

\begin{equation}
\label{eq:ttbar_enhance}
\mathcal{S} = \frac{\int_{m_{\ttbar} > 1\TeVcc} \frac{\rd\sigma_{SM +
      NP}} {\rd m_{\ttbar}}\rd m_{\ttbar} }{\int_{m_{\ttbar} > 1\TeVcc} \frac{\rd\sigma_{SM}}{\rd m_{\ttbar}}\rd m_{\ttbar}}.
\end{equation}

The events used for setting the limit
are selected to have reconstructed $m_{\ttbar} >$ 1\TeVcc,
which does not correspond to the same range for the true
mass. Consequently, a correction
factor must be applied to the reconstructed $\ttbar$ mass distribution
to estimate the true mass distribution. This is estimated by
dividing the number of simulated $\ttbar$ events with a reconstructed
mass $> 1$\TeVcc by the number of simulated $\ttbar$ events with a
true mass $> 1$\TeVcc.
This ratio is 1.24 $\pm$ 0.08
for the Type 1+1 analysis and 1.41 $\pm$ 0.11 for the Type 1+2
analysis.
These differences are applied as multiplicative factors to obtain the
yields for the true $\ttbar$ mass above 1\TeVcc.
These factors do not affect the quantity $\mathcal{S}$ since they cancel
in the ratio.

The approximate
NNLO cross section for inclusive
$\ttbar$ production is taken to be
163~pb~\cite{Aliev:2010zk,Langenfeld:2009wd,Kidonakis:2010dk}.
The efficiency for Type 1+1 events, relative to inclusive
SM $\ttbar$ production,
is found to be $(2.5 \pm 1.3)\times 10^{-4}$, and for
Type 1+2, the efficiency is $(1.6 \pm 1.0)\times
10^{-4}$.
The numbers of observed and expected events for the SM $\ttbar$ and NTMJ
backgrounds are shown
in Table~\ref{table:mttGt1TeV}, along with these efficiencies.
Following the statistical procedure outlined above, it follows that
the enhancement factor to the
$\ttbar$ production cross section for $m_{\ttbar} > 1$\TeVcc ($\mathcal{S}$ in
Eq.~(\ref{eq:ttbar_enhance})) must be $< 2.6$.
The a priori expectation is for this limit to lie in the interval
2.0--3.5 at 68\% CL, and 1.7--5.5 at 95\% CL, with a most probable value
of 2.5.

\begin{table}
\begin{center}
\caption{\label{table:mttGt1TeV}  Expected number of events with $m_{\ttbar} > 1$\TeVcc from SM $\ttbar$ and
  non-top multijet backgrounds, along with their total, compared to
  the observed number of events. The efficiency for SM $\ttbar$
  production, which is used in the limit setting procedure described
  in the text, is shown on the final line.}
\begin{tabular}{ |l|c|c| }
\hline
                             & 1+1     & 1+2   \\
\hline
Expected SM $\ttbar$ events      &  194 $\pm$ 106               &   129 $\pm$ 80      \\
Expected non-top multijet events & 1546 $\pm$ 45                &  2271 $\pm$ 130     \\
Total expected events            & 1740 $\pm$ 115               &  2400 $\pm$ 153     \\
Observed events                  & 1738                         &  2423               \\
\hline
$\ttbar$ efficiency              & $(2.5 \pm 1.3)\times 10^{-4}$ & $(1.6 \pm 1.0)\times 10^{-4}$      \\
\hline
\end{tabular}
\end{center}
\end{table}

\ifnpas
\section{Cross Checks}

\subsection{B-tagged Results}

As a cross check of the analysis, we present the $\ttbar$
invariant mass spectrum for the 1+2 analysis after requiring
at least one b-tag with the \emph{simple secondary vertex, high efficiency} tagging algorithm
with the medium operating point (\texttt{SSVHEM}).

It is rather difficult to estimate the b-tagging efficiency for these jets as this
has not yet been studied in detail. However, we overlay the Monte Carlo expectation
for the $\ttbar$ only, to show that we now have a significant top fraction in
this particular sample, and we do see that the data are basically consistent
with this so we are confident that we are identifying real boosted tops
with the top tagging and jet pruning algorithms.

\begin{figure}
\centering
\includegraphics[width=0.7\textwidth]{figs/TAGGED_TTMASS_TYPE12_JET300_TopMistagBkg}
\caption{1 + 2 analysis with $\ge 1$ b-tag (with the {\tt SSVHEM} b-tagger).
Only the Monte Carlo expectation from $\ttbar$ is shown along with the data.}
\label{figs:TAGGED_TTMASS_TYPE12_JET300_TopMistagBkg}
\end{figure}

In the semileptonic sample used to compute the subjet jet energy scale
and the substructure data-to-MC efficiency scale factor, many boosted
$\wboson$ jets are observed. However, no fully merged top jets are
observed (i.e., no type 1 hemispheres) due to small statistics.
As a check of observations of the type 1 hemispheres, an
additional $\mathrm{b}$ tag is required on the $\mathrm{b}$ jet
candidate in the 1 + 2
analysis. While the efficiency and background estimates
for this procedure have not been adequately determined at
the present time, from MC estimates a {\it true $\ttbar$}
purity of more than 60\% is expected. In this sample,
40 events are observed, of which a high fraction are predicted to be
boosted continuum $\ttbar$.
These events are tagged by three separate
algorithms: the top tagging algorithm, the $\wboson$ tagging
algorithm, and the $\mathrm{b}$ tagging algorithm.

One of these {\it golden} events is selected for graphical display.
Figure~\ref{figs:evDisplay_rhophi} shows a fully annotated $\rho-\phi$
view.
Figure~\ref{figs:evDisplay_rhophi_pf} shows the same event with the
particle flow candidates fully drawn, in the $\rho-\phi$ view,
and Figure~\ref{figs:evDisplay_lego_pf} shows the particle flow
candidates in the lego view.

\begin{figure}
\centering
\includegraphics[width=0.7\textwidth]{figs/goldenTopCand_1_white-166864_457688464_488_RhoPhi.png}
\caption{Event display of a golden triply-tagged 1 + 2
  candidate. The invariant mass of the $\ttbar$ candidate is 1352.5\GeVcc.
  In addition to the analysis selection, an additional
  $\mathrm{b}$ tagging requirement is made on the candidate
  $\mathrm{b}$ jet in the type 2 hemisphere. The type 1 top jet is shown in orange, with yellow
  denoting the three subjets. The type 2 hemisphere jets are shown
  in green. Jet 2 is tagged with the $\wboson$ tagging algorithm, and
  Jet 3 is tagged with a secondary vertex tag. The electromagnetic
  calorimeter information is shown in red, and the hadronic calorimieter
  information is shown in blue. }
\label{figs:evDisplay_rhophi}
\end{figure}

\begin{figure}
\centering
\includegraphics[width=0.7\textwidth]{figs/goldenTopCand_1_pfview_type2-166864_457688464_488_RhoPhi.pdf}
\caption{Event display of a golden triply-tagged 1 + 2
  candidate. In addition to the analysis selection, an additional
  $\mathrm{b}$ tagging requirement is made on the candidate
  $\mathrm{b}$ jet in the type 2 hemisphere. Here, the yellow corresponds to
  the particle flow candidates of the type 1 hemisphere jets, and
  the green corresponds to the particle flow candidates of the type
  2 hemisphere jets. The lines are charged and neutral particles.
  The electromagnetic
  calorimeter information is shown in red, and the hadronic calorimieter
  information is shown in blue.}
\label{figs:evDisplay_rhophi_pf}
\end{figure}

\begin{figure}
\centering
\includegraphics[width=0.7\textwidth]{figs/goldenTopCand_1_pfview_type2-166864_457688464_488_Lego.pdf}
\caption{Event display of a golden triply-tagged 1 + 2
  candidate. In addition to the analysis selection, an additional
  $\mathrm{b}$ tagging requirement is made on the candidate
  $\mathrm{b}$ jet in the type 2 hemisphere. Here, the yellow corresponds to
  the particle flow candidates of the type 1 hemisphere jets, and
  the green corresponds to the particle flow candidates of the type
  2 hemisphere jets. The height of the line is the energy measured
  by the particle flow algorithm for the various particles. The
  lines are charged and neutral particles. The electromagnetic
  calorimeter information is shown in red, and the hadronic calorimieter
  information is shown in blue.}
\label{figs:evDisplay_lego_pf}
\end{figure}

\fi

\section{Summary}
\ifnpas
\section{Conclusions}
\fi

\label{sec:conclusions}

In summary, a search is presented
for a massive resonance ($\zp$)
decaying into a $\ttbar$ pair in the all-hadronic final state using an integrated
luminosity of \intlumi\ collected with the CMS detector at 7\TeV.
A $\zp$ with standard-model couplings is considered as a model of such
a resonance. Two widths are considered
($\Gamma_{\zp} / \mzp = 1\%$ and 10\%), as well as an additional
model of
a Randall--Sundrum Kaluza--Klein gluon. The search focuses on
high $\ttbar$ masses that yield collimated decay products,
partially or fully merged into single jets.
The analysis therefore relies on new developments
in the area of jet substructure, thereby providing suppression of
non-top multijet production.

No excess of
events is observed over the expected yield from SM background sources.
Upper limits in the range of 1\unit{pb} are set on the product of
the $\zp$ cross section and branching fraction for a topcolor $\zp$ modeled
for several widths, as well as for a Randall--Sundrum Kaluza--Klein gluon.

Finally, results are presented for any generic source of new phenomena
with the same reconstruction efficiency as standard-model $\ttbar$ production, and limits
are placed on any enhancement to the cross section
from such a contribution. In particular,
the $\ttbar$ production cross section (in total) must be less than
a factor of 2.6 times that of the SM expectation for $m_{\ttbar} > 1$\TeVcc.
This constrains generic
enhancements to standard-model $\ttbar$ production, which can
be used to check models that seek to interpret the forward-backward
$\ttbar$ production asymmetry observed at the Tevatron
as a sign of new physics.

This is the first publication to constrain $\ttbar$ resonances in
the kinematic region of $m_{\ttbar} > 1$\TeVcc,
and is also the first work to use the jet-substructure tools
described above.

\section*{Acknowledgements}

We thank
Seung J. Lee for the computations of the Kaluza-Klein gluon
cross sections at 7 TeV.
We congratulate our colleagues in the CERN accelerator departments for
the excellent performance of the LHC machine. We thank the technical
and administrative staff at CERN and other CMS institutes, and
acknowledge support from: FMSR (Austria); FNRS and FWO (Belgium);
CNPq, CAPES, FAPERJ, and FAPESP (Brazil); MES (Bulgaria); CERN; CAS,
MoST, and NSFC (China); COLCIENCIAS (Colombia); MSES (Croatia); RPF
(Cyprus); MoER, SF0690030s09 and ERDF (Estonia); Academy of Finland,
MEC, and HIP (Finland); CEA and CNRS/IN2P3 (France); BMBF, DFG, and
HGF (Germany); GSRT (Greece); OTKA and NKTH (Hungary); DAE and DST
(India); IPM (Iran); SFI (Ireland); INFN (Italy); NRF and WCU (Korea);
LAS (Lithuania); CINVESTAV, CONACYT, SEP, and UASLP-FAI (Mexico); MSI
(New Zealand); PAEC (Pakistan); MSHE and NSC (Poland); FCT (Portugal);
JINR (Armenia, Belarus, Georgia, Ukraine, Uzbekistan); MON, RosAtom,
RAS and RFBR (Russia); MSTD (Serbia); MICINN and CPAN (Spain); Swiss
Funding Agencies (Switzerland); NSC (Taipei); TUBITAK and TAEK
(Turkey); STFC (United Kingdom); DOE and NSF (USA).

This work was supported in part by the DOE under under Task TeV
of contract DE-FG02-96ER40956 during the Workshop on Jet Substructure
at the University of Washington.

\clearpage

\bibliography{auto_generated}   

\providecommand{\href}[2]{#2}\begingroup\raggedright\begin{thebibliography}{10}%
\makeatletter
\providecommand{\hrefCMSnoop }[0]{\@secondoftwo}%
\makeatother
\providecommand{\doi}{\texttt{doi:}\begingroup \urlstyle{tt}\Url}

\bibitem{mssm}
\hrefCMSnoop {} {S.~Dimopoulos and H.~Georgi, ``{Softly Broken Supersymmetry
  and SU(5)}'',} \textit{ Nucl. Phys. B} \textbf{ 193} (1981) 150--162,
\href{http://dx.doi.org/10.1016/0550-3213(81)90522-8}{\doi{10.1016/0550-3213(81)90522-8}}.

\bibitem{nsd}
\hrefCMSnoop {} {S.~Weinberg, ``{Implications of Dynamical Symmetry
  Breaking}'',} \textit{ Phys. Rev. D} \textbf{ 13} (1976) 974--996,
\href{http://dx.doi.org/10.1103/PhysRevD.13.974}{\doi{10.1103/PhysRevD.13.974}}.

\bibitem{nsd2}
\hrefCMSnoop {} {L.~Susskind, ``{Dynamics of Spontaneous Symmetry Breaking in
  the Weinberg- Salam Theory}'',} \textit{ Phys. Rev. D} \textbf{ 20} (1979)
  2619--2625,
\href{http://dx.doi.org/10.1103/PhysRevD.20.2619}{\doi{10.1103/PhysRevD.20.2619}}.

\bibitem{nsd3}
\hrefCMSnoop {} {C.~T. Hill and S.~J. Parke, ``{Top production: Sensitivity to
  new physics}'',} \textit{ Phys. Rev. D} \textbf{ 49} (1994) 4454--4462,
  \href{http://dx.doi.org/10.1103/PhysRevD.49.4454}{\doi{10.1103/PhysRevD.49.4454}},
\href{http://www.arXiv.org/abs/hep-ph/9312324}{\texttt{ arXiv:hep-ph/9312324}}.

\bibitem{Hill:1991at}
\hrefCMSnoop {} {C.~T. Hill, ``{Topcolor: Top quark condensation in a gauge
  extension of the standard model}'',} \textit{ Phys. Lett. B} \textbf{ 266}
  (1991) 419--424,
  \href{http://dx.doi.org/10.1016/0370-2693(91)91061-Y}{\doi{10.1016/0370-2693(91)91061-Y}}.
Updates in
  \href{http://arxiv.org/abs/hep-ph/9911288}{\tt{arXiv:hep-ph/9911288}} and
  \href{http://arxiv.org/abs/1112.4928}{\tt{arXiv:hep-ph/1112.4928}}.

\bibitem{Hill:1994hp}
\hrefCMSnoop {} {C.~T. Hill, ``{Topcolor assisted technicolor}'',} \textit{
  Phys. Lett. B} \textbf{ 345} (1995) 483--489,
  \href{http://dx.doi.org/10.1016/0370-2693(94)01660-5}{\doi{10.1016/0370-2693(94)01660-5}},
\href{http://www.arXiv.org/abs/hep-ph/9411426}{\texttt{ arXiv:hep-ph/9411426}}.

\bibitem{nsd4}
\hrefCMSnoop {} {R.~S. Chivukula, B.~A. Dobrescu, H.~Georgi, and C.~T. Hill,
  ``{Top quark seesaw theory of electroweak symmetry breaking}'',} \textit{
  Phys. Rev. D} \textbf{ 59} (1999) 075003,
  \href{http://dx.doi.org/10.1103/PhysRevD.59.075003}{\doi{10.1103/PhysRevD.59.075003}},
\href{http://www.arXiv.org/abs/hep-ph/9809470}{\texttt{ arXiv:hep-ph/9809470}}.

\bibitem{littlehiggs}
\hrefCMSnoop {} {N.~Arkani-Hamed, A.~G. Cohen, and H.~Georgi, ``{Electroweak
  symmetry breaking from dimensional deconstruction}'',} \textit{ Phys. Lett.
  B} \textbf{ 513} (2001) 232--240,
  \href{http://dx.doi.org/10.1016/S0370-2693(01)00741-9}{\doi{10.1016/S0370-2693(01)00741-9}},
\href{http://www.arXiv.org/abs/hep-ph/0105239}{\texttt{ arXiv:hep-ph/0105239}}.

\bibitem{ed}
\hrefCMSnoop {} {N.~Arkani-Hamed, S.~Dimopoulos, and G.~R. Dvali, ``{The
  hierarchy problem and new dimensions at a millimeter}'',} \textit{ Phys.
  Lett. B} \textbf{ 429} (1998) 263--272,
  \href{http://dx.doi.org/10.1016/S0370-2693(98)00466-3}{\doi{10.1016/S0370-2693(98)00466-3}},
\href{http://www.arXiv.org/abs/hep-ph/9803315}{\texttt{ arXiv:hep-ph/9803315}}.

\bibitem{rs1}
\hrefCMSnoop {} {L.~Randall and R.~Sundrum, ``{A large mass hierarchy from a
  small extra dimension}'',} \textit{ Phys. Rev. Lett.} \textbf{ 83} (1999)
  3370--3373,
  \href{http://dx.doi.org/10.1103/PhysRevLett.83.3370}{\doi{10.1103/PhysRevLett.83.3370}},
\href{http://www.arXiv.org/abs/hep-ph/9905221}{\texttt{ arXiv:hep-ph/9905221}}.

\bibitem{rs2}
\hrefCMSnoop {} {L.~Randall and R.~Sundrum, ``{An alternative to
  compactification}'',} \textit{ Phys. Rev. Lett.} \textbf{ 83} (1999)
  4690--4693,
  \href{http://dx.doi.org/10.1103/PhysRevLett.83.4690}{\doi{10.1103/PhysRevLett.83.4690}},
\href{http://www.arXiv.org/abs/hep-th/9906064}{\texttt{ arXiv:hep-th/9906064}}.

\bibitem{rs_gluon_1}
K.~Agashe\hrefCMSnoop {} { {et~al.}, ``{LHC signals from warped extra
  dimensions}'',} \textit{ Phys. Rev. D} \textbf{ 77} (2008) 015003,
  \href{http://dx.doi.org/10.1103/PhysRevD.77.015003}{\doi{10.1103/PhysRevD.77.015003}},
\href{http://www.arXiv.org/abs/hep-ph/0612015}{\texttt{ arXiv:hep-ph/0612015}}.

\bibitem{bhkr}
\hrefCMSnoop {} {Y.~Bai, J.~L. Hewett, J.~Kaplan, and T.~G. Rizzo, ``{LHC
  Predictions from a Tevatron Anomaly in the Top Quark Forward-Backward
  Asymmetry}'',} \textit{ JHEP} \textbf{ 03} (2011) 003,
  \href{http://dx.doi.org/10.1007/JHEP03(2011)003}{\doi{10.1007/JHEP03(2011)003}},
  \href{http://www.arXiv.org/abs/1101.5203}{\texttt{ arXiv:1101.5203}}.

\bibitem{axigluon_ttbar_1}
\hrefCMSnoop {} {P.~H. Frampton, J.~Shu, and K.~Wang, ``{Axigluon as Possible
  Explanation for $p\bar{p} \to t\bar{t}$ Forward-Backward Asymmetry}'',}
  \textit{ Phys. Lett. B} \textbf{ 683} (2010) 294--297,
  \href{http://dx.doi.org/10.1016/j.physletb.2009.12.043}{\doi{10.1016/j.physletb.2009.12.043}},
\href{http://www.arXiv.org/abs/0911.2955}{\texttt{ arXiv:0911.2955}}.

\bibitem{axigluon_ttbar_2}
\hrefCMSnoop {} {M.~I. Gresham, I.-W. Kim, and K.~M. Zurek, ``{On Models of New
  Physics for the Tevatron Top A\_FB}'',} \textit{ Phys. Rev. D} \textbf{ 83}
  (2011) 114027,
  \href{http://dx.doi.org/10.1103/PhysRevD.83.114027}{\doi{10.1103/PhysRevD.83.114027}},
\href{http://www.arXiv.org/abs/1103.3501}{\texttt{ arXiv:1103.3501}}.

\bibitem{axigluon_ttbar_3}
\hrefCMSnoop {} {O.~Antunano, J.~H. Kuhn, and G.~Rodrigo, ``{Top quarks,
  axigluons and charge asymmetries at hadron colliders}'',} \textit{ Phys. Rev.
  D} \textbf{ 77} (2008) 014003,
  \href{http://dx.doi.org/10.1103/PhysRevD.77.014003}{\doi{10.1103/PhysRevD.77.014003}},
\href{http://www.arXiv.org/abs/0709.1652}{\texttt{ arXiv:0709.1652}}.

\bibitem{top_AFB1}
\hrefCMSnoop {} {{ CDF} Collaboration, ``{Forward-Backward Asymmetry in Top
  Quark Production in $\mathrm{p}\overline{\mathrm{p}}$ Collisions at
  $\sqrt{s}$ = 1.96~TeV}'',} \textit{ Phys. Rev. Lett.} \textbf{ 101} (2008)
  202001,
  \href{http://dx.doi.org/10.1103/PhysRevLett.101.202001}{\doi{10.1103/PhysRevLett.101.202001}},
\href{http://www.arXiv.org/abs/0806.2472}{\texttt{ arXiv:0806.2472}}.

\bibitem{top_AFB2}
\hrefCMSnoop {} {{ D0} Collaboration, ``{First measurement of the
  forward-backward charge asymmetry in top quark pair production}'',} \textit{
  Phys. Rev. Lett.} \textbf{ 100} (2008) 142002,
  \href{http://dx.doi.org/10.1103/PhysRevLett.100.142002}{\doi{10.1103/PhysRevLett.100.142002}},
\href{http://www.arXiv.org/abs/0712.0851}{\texttt{ arXiv:0712.0851}}.

\bibitem{top_AFB3}
\hrefCMSnoop {} {{ CDF} Collaboration, ``{Evidence for a Mass Dependent
  Forward-Backward Asymmetry in Top Quark Pair Production}'',} \textit{ Phys.
  Rev. D} \textbf{ 83} (2011) 112003,
  \href{http://dx.doi.org/10.1103/PhysRevD.83.112003}{\doi{10.1103/PhysRevD.83.112003}},
\href{http://www.arXiv.org/abs/1101.0034}{\texttt{ arXiv:1101.0034}}.

\bibitem{top_AFB4}
\hrefCMSnoop {} {{ D0} Collaboration, ``{Forward-backward asymmetry in top
  quark-antiquark production}'',} \textit{ Phys. Rev. D} \textbf{ 84} (2011)
  112005,
  \href{http://dx.doi.org/10.1103/PhysRevD.84.112005}{\doi{10.1103/PhysRevD.84.112005}},
\href{http://www.arXiv.org/abs/1107.4995}{\texttt{ arXiv:1107.4995}}.

\bibitem{top_AFB5}
\hrefCMSnoop {} {{ CDF} Collaboration, ``{Evidence for a Mass Dependent
  Forward-Backward Asymmetry in Top Quark Pair Production}'',} \textit{ Phys.
  Rev. D} \textbf{ 83} (2011) 112003,
  \href{http://dx.doi.org/10.1103/PhysRevD.83.112003}{\doi{10.1103/PhysRevD.83.112003}},
\href{http://www.arXiv.org/abs/1101.0034}{\texttt{ arXiv:1101.0034}}.

\bibitem{top_afb_implications1}
C.~Delaunay\hrefCMSnoop {} { {et~al.}, ``{Implications of the CDF $t \bar{t}$
  Forward-Backward Asymmetry for Hard Top Physics}'',} \textit{ JHEP} \textbf{
  08} (2011) 031,
  \href{http://dx.doi.org/10.1007/JHEP08(2011)031}{\doi{10.1007/JHEP08(2011)031}},
  \href{http://www.arXiv.org/abs/1103.2297}{\texttt{ arXiv:1103.2297}}.

\bibitem{top_afb_implications2}
\hrefCMSnoop {} {J.~A. Aguilar-Saavedra and M.~Perez-Victoria, ``{Probing the
  Tevatron $\mathrm{t}\overline{\mathrm{t}}$ asymmetry at LHC}'',} \textit{
  JHEP} \textbf{ 05} (2011) 034,
  \href{http://dx.doi.org/10.1007/JHEP05(2011)034}{\doi{10.1007/JHEP05(2011)034}},
  \href{http://www.arXiv.org/abs/1103.2765}{\texttt{ arXiv:1103.2765}}.

\bibitem{cdftt1}
\hrefCMSnoop {} {{ CDF} Collaboration, ``{Limits on the production of narrow
  $\mathrm{t}\overline{\mathrm{t}}$ resonances in
  $\mathrm{p}\overline{\mathrm{p}}$ collisions at $\sqrt{s} =$ 1.96~TeV}'',}
  \textit{ Phys. Rev. D} \textbf{ 77} (2008) 051102,
  \href{http://dx.doi.org/10.1103/PhysRevD.77.051102}{\doi{10.1103/PhysRevD.77.051102}},
\href{http://www.arXiv.org/abs/0710.5335}{\texttt{ arXiv:0710.5335}}.

\bibitem{cdftt2}
\hrefCMSnoop {} {{ CDF} Collaboration, ``{Search for resonant
  $\mathrm{t}\overline{\mathrm{t}}$ production in
  $\mathrm{p}\overline{\mathrm{p}}$ collisions at $\sqrt{s}$ = 1.96~TeV}'',}
  \textit{ Phys. Rev. Lett.} \textbf{ 100} (2008) 231801,
  \href{http://dx.doi.org/10.1103/PhysRevLett.100.231801}{\doi{10.1103/PhysRevLett.100.231801}},
\href{http://www.arXiv.org/abs/0709.0705}{\texttt{ arXiv:0709.0705}}.

\bibitem{d0tt}
\hrefCMSnoop {} {{ D0} Collaboration, ``{Search for a Narrow $t\bar{t}$
  Resonance in $p\bar{p}$ Collisions at $\sqrt{s}=1.96$ TeV}'',} \textit{ Phys.
  Rev. D} \textbf{ 85} (2012) 051101,
  \href{http://dx.doi.org/10.1103/PhysRevD.85.051101}{\doi{10.1103/PhysRevD.85.051101}},
\href{http://www.arXiv.org/abs/1111.1271}{\texttt{ arXiv:1111.1271}}.

\bibitem{catop_theory}
\hrefCMSnoop {} {D.~E. Kaplan, K.~Rehermann, M.~D. Schwartz, and B.~Tweedie,
  ``{Top Tagging: A Method for Identifying Boosted Hadronically Decaying Top
  Quarks}'',} \textit{ Phys. Rev. Lett.} \textbf{ 101} (2008) 142001,
  \href{http://dx.doi.org/10.1103/PhysRevLett.101.142001}{\doi{10.1103/PhysRevLett.101.142001}},
\href{http://www.arXiv.org/abs/0806.0848}{\texttt{ arXiv:0806.0848}}.

\bibitem{catop_cms}
\href {https://cdsweb.cern.ch/record/1194489} {{ CMS} Collaboration, ``A
  Cambridge-Aachen (C-A) based Jet Algorithm for boosted top-jet tagging'',}
  CMS Physics Analysis Summary CMS-PAS-JME-009-01, (2009).

\bibitem{jetpruning1}
\hrefCMSnoop {} {S.~D. Ellis, C.~K. Vermilion, and J.~R. Walsh, ``{Techniques
  for improved heavy particle searches with jet substructure}'',} \textit{
  Phys. Rev. D} \textbf{ 80} (2009) 051501,
  \href{http://dx.doi.org/10.1103/PhysRevD.80.051501}{\doi{10.1103/PhysRevD.80.051501}},
  \href{http://www.arXiv.org/abs/0903.5081}{\texttt{ arXiv:0903.5081}}.

\bibitem{jetpruning2}
\hrefCMSnoop {} {S.~D. Ellis, C.~K. Vermilion, and J.~R. Walsh,
  ``{Recombination Algorithms and Jet Substructure: Pruning as a Tool for Heavy
  Particle Searches}'',} \textit{ Phys. Rev. D} \textbf{ 81} (2010) 094023,
  \href{http://dx.doi.org/10.1103/PhysRevD.81.094023}{\doi{10.1103/PhysRevD.81.094023}},
  \href{http://www.arXiv.org/abs/0912.0033}{\texttt{ arXiv:0912.0033}}.

\bibitem{boostedhiggs}
\hrefCMSnoop {} {J.~M. Butterworth, A.~R. Davison, M.~Rubin, and G.~P. Salam,
  ``{Jet substructure as a new Higgs search channel at the LHC}'',} \textit{
  Phys. Rev. Lett.} \textbf{ 100} (2008) 242001,
  \href{http://dx.doi.org/10.1103/PhysRevLett.100.242001}{\doi{10.1103/PhysRevLett.100.242001}},
\href{http://www.arXiv.org/abs/0802.2470}{\texttt{ arXiv:0802.2470}}.

\bibitem{:2008zzk}
\hrefCMSnoop {} {{ CMS} Collaboration, ``{The CMS experiment at the CERN
  LHC}'',} \textit{ JINST} \textbf{ 3} (2008) S08004,
\href{http://dx.doi.org/10.1088/1748-0221/3/08/S08004}{\doi{10.1088/1748-0221/3/08/S08004}}.

\bibitem{MadGraph}
J.~Alwall\hrefCMSnoop {} { {et~al.}, ``{MadGraph/MadEvent v4: the new web
  generation}'',} \textit{ JHEP} \textbf{ 09} (2007) 028,
  \href{http://dx.doi.org/10.1088/1126-6708/2007/09/028}{\doi{10.1088/1126-6708/2007/09/028}},
\href{http://www.arXiv.org/abs/0706.2334}{\texttt{ arXiv:0706.2334}}.

\bibitem{pythia}
T.~Sj{\"o}strand\hrefCMSnoop {} { {et~al.}, ``{High-energy physics event
  generation with PYTHIA 6.1}'',} \textit{ Comput. Phys. Commun.} \textbf{ 135}
  (2001) 238--259,
  \href{http://dx.doi.org/10.1016/S0010-4655(00)00236-8}{\doi{10.1016/S0010-4655(00)00236-8}},
\href{http://www.arXiv.org/abs/hep-ph/0010017}{\texttt{ arXiv:hep-ph/0010017}}.

\bibitem{pythia8}
\hrefCMSnoop {} {T.~Sjostrand, S.~Mrenna, and P.~Z. Skands, ``{A Brief
  Introduction to PYTHIA 8.1}'',} \textit{ Comput. Phys. Commun.} \textbf{ 178}
  (2008) 852--867,
  \href{http://dx.doi.org/10.1016/j.cpc.2008.01.036}{\doi{10.1016/j.cpc.2008.01.036}},
\href{http://www.arXiv.org/abs/0710.3820}{\texttt{ arXiv:0710.3820}}.

\bibitem{cteq}
J.~Pumplin\hrefCMSnoop {} { {et~al.}, ``{New generation of parton distributions
  with uncertainties from global QCD analysis}'',} \textit{ JHEP} \textbf{ 07}
  (2002) 012,
  \href{http://dx.doi.org/10.1088/1126-6708/2002/07/012}{\doi{10.1088/1126-6708/2002/07/012}},
  \href{http://www.arXiv.org/abs/hep-ph/0201195}{\texttt{
  arXiv:hep-ph/0201195}}.

\bibitem{Geant4}
\hrefCMSnoop {} {J.~Allison {et~al.}, ``{Geant4} developments and
  applications'',} \textit{ IEEE Trans. Nucl. Sci.} \textbf{ 53} (2006) 270,
\href{http://dx.doi.org/10.1109/TNS.2006.869826}{\doi{10.1109/TNS.2006.869826}}.

\bibitem{particleflow}
\href {https://cdsweb.cern.ch/record/1194487} {{ CMS} Collaboration,
  ``{Particle-Flow Event Reconstruction in CMS and Performance for Jets, Taus,
  and MET}'',} CMS Physics Analysis Summary CMS-PAS-PFT-09-01, (2009).

\bibitem{highpuritytracks}
\href {https://cdsweb.cern.ch/record/1258204} {{ CMS} Collaboration, ``Tracking
  and Vertexing Results from First Collisions'',} CMS Physics Analysis Summary
  CMS-PAS-TRK-10-001, (2010).

\bibitem{CAaachen}
\hrefCMSnoop {} {M.~Wobisch and T.~Wengler, ``{Hadronization corrections to jet
  cross sections in deep- inelastic scattering}'',} \textit{ DESY-PROC}
  \textbf{ 02} (1998)
\href{http://www.arXiv.org/abs/hep-ph/9907280}{\texttt{ arXiv:hep-ph/9907280}}.

\bibitem{CAcambridge}
\hrefCMSnoop {} {Y.~L. Dokshitzer, G.~D. Leder, S.~Moretti, and B.~R. Webber,
  ``{Better Jet Clustering Algorithms}'',} \textit{ JHEP} \textbf{ 08} (1997)
  001,
  \href{http://dx.doi.org/10.1088/1126-6708/1997/08/001}{\doi{10.1088/1126-6708/1997/08/001}},
\href{http://www.arXiv.org/abs/hep-ph/9707323}{\texttt{ arXiv:hep-ph/9707323}}.

\bibitem{fastjet1}
\hrefCMSnoop {} {M.~Cacciari and G.~P. Salam, ``{Dispelling the N**3 myth for
  the k(t) jet-finder}'',} \textit{ Phys. Lett. B} \textbf{ 641} (2006) 57--61,
  \href{http://dx.doi.org/10.1016/j.physletb.2006.08.037}{\doi{10.1016/j.physletb.2006.08.037}},
\href{http://www.arXiv.org/abs/hep-ph/0512210}{\texttt{ arXiv:hep-ph/0512210}}.

\bibitem{fastjet2}
\hrefCMSnoop {} {M.~Cacciari, G.~P. Salam, and G.~Soyez, ``{FastJet user
  manual}'',} (2011).
\href{http://www.arXiv.org/abs/1111.6097}{\texttt{ arXiv:1111.6097}}.

\bibitem{ktalg}
\hrefCMSnoop {} {M.~Cacciari, G.~P. Salam, and G.~Soyez, ``{The anti-kt jet
  clustering algorithm}'',} \textit{ JHEP} \textbf{ 04} (2008) 063,
\href{http://www.arXiv.org/abs/0802.1189}{\texttt{ arXiv:0802.1189}}.

\bibitem{jec_jinst}
\hrefCMSnoop {} {{ CMS} Collaboration, ``{Determination of Jet Energy
  Calibration and Transverse Momentum Resolution in CMS}'',} \textit{ JINST}
  \textbf{ 6} (2011) P11002,
  \href{http://dx.doi.org/10.1088/1748-0221/6/11/P11002}{\doi{10.1088/1748-0221/6/11/P11002}},
\href{http://www.arXiv.org/abs/1107.4277}{\texttt{ arXiv:1107.4277}}.

\bibitem{TOP-10-003}
\hrefCMSnoop {} {{ CMS} Collaboration, ``{Measurement of the t $\bar{t}$
  Production Cross Section in pp Collisions at 7 TeV in Lepton + Jets Events
  Using b-quark Jet Identification}'',} \textit{ Phys. Rev. D} \textbf{ 84}
  (2011) 092004,
  \href{http://dx.doi.org/10.1103/PhysRevD.84.092004}{\doi{10.1103/PhysRevD.84.092004}},
\href{http://www.arXiv.org/abs/1108.3773}{\texttt{ arXiv:1108.3773}}.

\bibitem{CMS-PAS-BTV-11-001}
\href {http://cdsweb.cern.ch/record/1366061/} {{ CMS} Collaboration,
  ``Performance of the b-jet identification in CMS'',} CMS Physics Analysis
  Summary CMS-PAS-BTV-11-001, (2011).

\bibitem{Aliev:2010zk}
\hrefCMSnoop {} {M.~Aliev {et~al.}, ``{HATHOR: HAdronic Top and Heavy quarks
  crOss section calculatoR}'',} \textit{ Comput. Phys. Commun.} \textbf{ 182}
  (2011) 1034,
  \href{http://dx.doi.org/10.1016/j.cpc.2010.12.040}{\doi{10.1016/j.cpc.2010.12.040}},
  \href{http://www.arXiv.org/abs/1007.1327}{\texttt{ arXiv:1007.1327}}.

\bibitem{Langenfeld:2009wd}
\hrefCMSnoop {} {U.~Langenfeld, S.~Moch, and P.~Uwer, ``{Measuring the running
  top-quark mass}'',} \textit{ Phys. Rev. D} \textbf{ 80} (2009) 054009,
  \href{http://dx.doi.org/10.1103/PhysRevD.80.054009}{\doi{10.1103/PhysRevD.80.054009}},
  \href{http://www.arXiv.org/abs/0906.5273}{\texttt{ arXiv:0906.5273}}.

\bibitem{Kidonakis:2010dk}
\hrefCMSnoop {} {N.~Kidonakis, ``{Next-to-next-to-leading soft-gluon
  corrections for the top quark cross section and transverse momentum
  distribution}'',} \textit{ Phys. Rev. D} \textbf{ 82} (2010) 114030,
  \href{http://dx.doi.org/10.1103/PhysRevD.82.114030}{\doi{10.1103/PhysRevD.82.114030}},
  \href{http://www.arXiv.org/abs/1009.4935}{\texttt{ arXiv:1009.4935}}.

\bibitem{fewz}
\hrefCMSnoop {} {K.~Melnikov and F.~Petriello, ``Electroweak gauge boson
  production at hadron colliders through $O(\alpha_s^2)$'',} \textit{ Phys.
  Rev. D} \textbf{ 74} (2006) 114017,
  \href{http://dx.doi.org/10.1103/PhysRevD.74.114017}{\doi{10.1103/PhysRevD.74.114017}},
\href{http://www.arXiv.org/abs/hep-ph/0609070}{\texttt{ arXiv:hep-ph/0609070}}.

\bibitem{Ellis:2007ib}
S.~D. Ellis\hrefCMSnoop {} { {et~al.}, ``{Jets in hadron-hadron collisions}'',}
  \textit{ Prog. Part. Nucl. Phys.} \textbf{ 60} (2008) 484--551,
  \href{http://dx.doi.org/10.1016/j.ppnp.2007.12.002}{\doi{10.1016/j.ppnp.2007.12.002}},
\href{http://www.arXiv.org/abs/0712.2447}{\texttt{ arXiv:0712.2447}}.

\bibitem{pdf4lhc}
\hrefCMSnoop {} {G.~Watt, ``{Parton distribution function dependence of
  benchmark Standard Model total cross sections at the 7 TeV LHC}'',} \textit{
  JHEP} \textbf{ 09} (2011) 069,
  \href{http://dx.doi.org/10.1007/JHEP09(2011)069}{\doi{10.1007/JHEP09(2011)069}},
  \href{http://www.arXiv.org/abs/1106.5788}{\texttt{ arXiv:1106.5788}}.

\bibitem{Read1}
\hrefCMSnoop {} {A.~L. Read, ``Presentation of search results: the {CLs}
  technique'',} \textit{ J. Phys. G} \textbf{ 28} (2002) 2693,
  \href{http://dx.doi.org/10.1088/0954-3899/28/10/313}{\doi{10.1088/0954-3899/28/10/313}}.

\bibitem{junkcls}
\hrefCMSnoop {} {T.~Junk, ``{Confidence level computation for combining
  searches with small statistics}'',} \textit{ Nucl. Instrum. Meth. A} \textbf{
  434} (1999) 435,
  \href{http://dx.doi.org/10.1016/S0168-9002(99)00498-2}{\doi{10.1016/S0168-9002(99)00498-2}}.

\bibitem{cls_lhc}
\href {https://cdsweb.cern.ch/record/1379837} {{ATLAS and CMS Collaborations},
  ``Procedure for the LHC Higgs boson search combination in summer 2011'',}
  ATL-PHYS-PUB-2011-011, {CMS NOTE-11-005}, (2011).

\bibitem{Harris:2011ez}
\hrefCMSnoop {} {R.~M. Harris and S.~Jain, ``{Cross Sections for Leptophobic
  Topcolor Z' decaying to top-antitop}'',} \textit{ Eur. Phys. J. C} \textbf{
  72} (2012) 2072,
\href{http://www.arXiv.org/abs/1112.4928}{\texttt{ arXiv:1112.4928}}.

\end{thebibliography}\endgroup

\appendix

\cleardoublepage \appendix\section{The CMS Collaboration \label{app:collab}}\begin{sloppypar}\hyphenpenalty=5000\widowpenalty=500\clubpenalty=5000\textbf{Yerevan Physics Institute,  Yerevan,  Armenia}\\*[0pt]
S.~Chatrchyan, V.~Khachatryan, A.M.~Sirunyan, A.~Tumasyan
\vskip\cmsinstskip
\textbf{Institut f\"{u}r Hochenergiephysik der OeAW,  Wien,  Austria}\\*[0pt]
W.~Adam, T.~Bergauer, M.~Dragicevic, J.~Er\"{o}, C.~Fabjan, M.~Friedl, R.~Fr\"{u}hwirth, V.M.~Ghete, J.~Hammer\cmsAuthorMark{1}, N.~H\"{o}rmann, J.~Hrubec, M.~Jeitler, W.~Kiesenhofer, M.~Krammer, D.~Liko, I.~Mikulec, M.~Pernicka$^{\textrm{\dag}}$, B.~Rahbaran, C.~Rohringer, H.~Rohringer, R.~Sch\"{o}fbeck, J.~Strauss, A.~Taurok, F.~Teischinger, P.~Wagner, W.~Waltenberger, G.~Walzel, E.~Widl, C.-E.~Wulz
% ...
\vskip\cmsinstskip
\dag:~Deceased\\
1:~~Also at CERN, European Organization for Nuclear Research, Geneva, Switzerland\\
% ...

\end{sloppypar}
\end{document}